%% file: 0paper.tex
\documentclass[sigconf,screen,nonacm]{acmart}
\input{00packages.tex}

\AtBeginDocument{%
  \providecommand\BibTeX{{%
    \normalfont B\kern-0.5em{\scshape i\kern-0.25em b}\kern-0.8em\TeX}}}

\begin{document}


\title[Mental Health Impacts of AI Companions]{Mental Health Impacts of AI Companions}
\subtitle{Triangulating Social Media Quasi-Experiments, User Perspectives, and Relational Theory}

\author{Yunhao Yuan}
\orcid{0000-0002-1450-8572}
\affiliation{%
  \institution{Aalto University}
  \city{Espoo}
  \state{}
  \country{Finland}
}
\email{yunhao.yuan@aalto.fi}

\author{Jiaxun Zhang}
\orcid{0009-0004-1054-4359}
\affiliation{%
  \institution{University of Illinois Urbana-Champaign}
  \city{Urbana}
  \state{Illinois}
  \country{USA}}
\email{jiaxunz2@illinois.edu}

\author{Talayeh Aledavood}
\orcid{0000-0002-0110-5694}
\affiliation{%
  \institution{Aalto University}
  \city{Espoo}
  \state{}
  \country{Finland}}
\email{talayeh.aledavood@aalto.fi}

\author{Renwen Zhang}
\orcid{0000-0002-7636-9598}
\affiliation{
  \institution{Nanyang Technological University}
  \city{Singapore}
  \state{}
  \country{Singapore}}
\email{renwen.zhang@ntu.edu.sg}

\author{Koustuv Saha}
\orcid{0000-0002-8872-2934}
\affiliation{%
  \institution{University of Illinois Urbana-Champaign}
  \city{Urbana}
  \state{IL}
  \country{USA}}
\email{ksaha2@illinois.edu}


\renewcommand{\shortauthors}{Yunhao Yuan et al.}

\input{0abstract}

\ccsdesc[300]{Human-centered computing~Empirical studies in collaborative and social computing}
\ccsdesc[300]{Applied computing~Psychology}
\ccsdesc[300]{Human-centered computing~Social media}

\keywords{AI companions, mental health, wellbeing, causal inference, Reddit, language, Replika, relationship, loneliness}

\maketitle


\input{1introduction.tex} 
\input{2relatedwork.tex} 
\input{3data.tex} 

\input{5_1results_quantitative} 
\input{5_2results_qualitative_v2} 
\input{6discussion} 
\input{7limitations.tex} 
\input{8conclusions.tex} 


\begin{acks}
We acknowledge the computational resources provided by the Aalto
Science-IT project. This research was largely conducted during Yuan's visiting researcher appointment at the University of Illinois Urbana–Champaign, supported by funding from the Helsinki Institute for Information Technology (HIIT). Renwen Zhang is supported by the Nanyang Technological University Start-up Grant (NAP\_SUG 025564-00001). 

\end{acks}

\bibliographystyle{ACM-Reference-Format}
\bibliography{0paper}

%
\appendix
\input{9.appendix_parallel_trends}

\end{document}
\endinput



%% file: 00packages.tex
\usepackage{color, colortbl, xcolor}
\usepackage{url}
\usepackage{subcaption}
\usepackage{textcomp}
\usepackage{soul}
\usepackage{multirow}
\usepackage{enumitem}
\usepackage{mathtools}
\usepackage{siunitx}
\usepackage{array}
\usepackage{colortbl}
\usepackage{hhline}

\usepackage{booktabs} 
\usepackage{array}
\usepackage{xcolor}

\newcommand*{\rowstyle}[1]{
  \gdef\@rowstyle{#1}%
  \@rowstyle\ignorespaces%
}

\newcolumntype{=}{
  >{\gdef\@rowstyle{}}%
}

\newcolumntype{+}{
  >{\@rowstyle}%
}

\usepackage{arydshln}
\definecolor{linkColor}{RGB}{6,125,233}
\definecolor{green}{rgb}{0.0, 0.65, 0.31}
\definecolor{bleudefrance}{rgb}{0.19, 0.55, 0.91}
\definecolor{ceruleanblue}{rgb}{0.16, 0.32, 0.75}
\definecolor{grey}{HTML}{969696}
\definecolor{violet}{HTML}{756bb1}
\definecolor{dgrey}{HTML}{01665e}
\definecolor{lgrey}{HTML}{5ab4ac}
\definecolor{dgreen}{HTML}{005a32}
\definecolor{purple}{HTML}{54278f}

\definecolor{editCol}{HTML}{000000}
\definecolor{maskCol}{HTML}{c51b7d}
\definecolor{lrColor}{HTML}{8856a7}
\definecolor{trColor}{HTML}{d01c8b}
\definecolor{ctColor}{HTML}{4dac26}
\definecolor{brickred}{HTML}{f03b20}
\definecolor{improveCol}{HTML}{253494}
\definecolor{worsenCol}{HTML}{d7191c}
\definecolor{DarkBlue}{HTML}{00008B}
\definecolor{mscolor}{HTML}{01665e}
\definecolor{nmscolor}{HTML}{bf812d}
\definecolor{lgreen}{HTML}{ccece6}
\definecolor{dolive}{HTML}{308014}
\definecolor{dpink}{HTML}{CD1076}
\definecolor{soothinggreen}{HTML}{4dac26}
\definecolor{darkred}{HTML}{8B0000}

\colorlet{tablerowcolor4}{gray!50} 

\newcommand*{\textlabel}[2]{%
  \edef\@currentlabel{#1}
  \phantomsection
  #1\label{#2}
}

\colorlet{tableheadcolor}{gray!25} 
\colorlet{tablerowcolor}{gray!10} 
\colorlet{tablerowcolor2}{gray!45} 
\colorlet{tablerowcolor3}{gray!25} 

\newcommand{\rowcol}{\rowcolor{tablerowcolor}} %

\newcolumntype{a}{>{\columncolor{tablerowcolor}}r}
\definecolor{aicolor}{HTML}{018571}
\definecolor{occolor}{HTML}{ff7799}

\definecolor{aicolor}{HTML}{fc8d62}
\definecolor{occolor}{HTML}{253494}

\newif{\ifhidecomments}
 \hidecommentsfalse 
\ifhidecomments
    \newcommand{\yunhao}[1]{}
    \newcommand{\jiaxun}[1]{}
    \newcommand{\renwen}[1]{}
    \newcommand{\talayeh}[1]{}
    \newcommand{\koustuv}[1]{}
\else
    \newcommand{\yunhao}[1]{\textbf{\small\sffamily{\textcolor{DarkBlue}{[#1 -- Yunhao]}}}}
    \newcommand{\jiaxun}[1]{\textbf{\small\sffamily{\textcolor{orange}{[#1 -- Jiaxun]}}}}
    \newcommand{\talayeh}[1]{\textbf{\small\sffamily{\textcolor{violet}{[#1 -- Talayeh]}}}}
    \newcommand{\renwen}[1]{\textbf{\small\sffamily{\textcolor{dgreen}{[#1 -- Renwen]}}}}
    \newcommand{\koustuv}[1]{\textbf{\small\sffamily{\textcolor{dpink}{[#1 -- Koustuv]}}}}
  \fi
\newcommand{\edit}[1]{{\textcolor{editCol}{#1}}}

\newcommand{\Tr}{\textsf{Treatment}}
\newcommand{\Cto}{\textsf{Control-VA}}
\newcommand{\Ctt}{\textsf{Control-NAI}}
\newcommand{\Cllm}{\textsf{Control-LLM}}

\newcommand{\Ct}{\textsf{Control}}
\newcommand{\acc}{AICC}
\newcommand{\n}[1]{$\mathtt{#1}$}

\renewcommand{\textrightarrow}{$\rightarrow$}

\colorlet{tableheadcolor}{gray!25} 

\definecolor{neutralCol}{HTML}{dd1c77}
\definecolor{neutralGreen}{HTML}{31a354}
\definecolor{NewBlue}{HTML}{1879ba}
\definecolor{bleudefrance}{rgb}{0.19, 0.55, 0.91}  
\definecolor{AfTrColor}{HTML}{0868ac}  
\definecolor{BfTrColor}{HTML}{a8ddb5}  

\definecolor{AfCtColor}{HTML}{b10026}  
\definecolor{BfCtColor}{HTML}{fd8d3c}

\graphicspath{ {figures/} }

\newcommand{\para}[1]{\vspace{0.3em}\noindent\textbf{#1}~}

\newcolumntype{C}[1]{>{\centering\arraybackslash}p{#1}}

%% file: 0abstract.tex
\begin{abstract}
AI-powered companion chatbots (\acc{}s) such as Replika are increasingly popular, offering empathetic interactions, yet their psychosocial impacts remain unclear. We examined how engaging with \acc{}s shaped wellbeing and how users perceived these experiences. First, we conducted a large-scale quasi-experimental study of longitudinal Reddit data, applying stratified propensity score matching and Difference-in-Differences regression. 
Findings revealed mixed effects---greater grief expression and interpersonal focus, alongside increases in language about loneliness, depression, and suicidal ideation. Second, we complemented these results with \edit{18} semi-structured interviews, which we thematically analyzed and contextualized using Knapp’s relationship development model. We identified trajectories of initiation, escalation, and bonding, wherein \acc{}s provided emotional validation and social rehearsal but also carried risks of over-reliance and withdrawal. Triangulating across methods, we offer design implications for AI companions that scaffold healthy boundaries, support mindful engagement, support disclosure without dependency, and surface relationship stages—maximizing psychosocial benefits while mitigating risks.


\end{abstract}

%% file: 1introduction.tex
\section{Introduction}

Advances in generative AI (genAI) and large language models (LLMs) have driven the rise of AI-powered companion chatbots (\acc{}), such as Replika~\cite{wikipediaReplikaWikipedia} and Character.AI~\cite{wikipediaCharacterAIWikipedia}, which have gained millions of users worldwide. 
\acc{}s go beyond task-oriented chatbots by offering personalized, interactive conversations that often emulate empathetic social interactions. 
\acc{}s are marketed as accessible and non-judgmental companions which provide emotional support, companionship, and guidance in daily life~\cite{skjuve2021my}.
This promise of mitigating loneliness and fostering a sense of connection has contributed to their growing popularity. 
However, despite this potential, the growing adoption of \acc{}s raises critical questions about their broader impact. 

Recent research has identified several promising outcomes associated with using \acc{}s, including reduced feelings of loneliness, improved emotional coping, and potential contributions to suicide prevention~\cite{de2024ai,syed2024role,maples2024loneliness}. At the same time, concerns have emerged around issues such as emotional overdependence, heightened vulnerability, altered social practices, problematic or harmful responses, and risks tied to sensitive interactions~\cite{zhang2025rise,zhang2024dark}.
Much of the existing scholarship on \acc{} draws on qualitative insights and self-reported experiences, which have been critical to understanding user perceptions and relationships. 
However, these approaches often capture short-term, static perspectives and are subject to inherent limitations such as recall bias and limited sample diversity~\cite{tourangeau2000psychology}. Consequently, little is known about the longer-term psychosocial consequences of sustained engagement with \acc{}s.

Addressing this knowledge gap requires methodological innovation. 
Prior work has largely relied on descriptive accounts and self-reports~\cite{skjuve2021my,skjuve2022longitudinal,alabed2024more,brandtzaeg2022my}, whereas online behavioral traces provide a powerful complementary lens~\cite{de2016discovering,kiciman2018using,saha2019social}.
Social media platforms, in particular, enable individuals to candidly self-disclose in a self-motivated fashion~\cite{de2014mental}, producing both linguistic (and psycholinguistic) signals and behavioral signals (e.g., activity patterns, engagement) that together offer a powerful lens on psychosocial processes. 
Such data has enabled large-scale studies on mental health~\cite{de2013predicting}, and prior research has established the construct validity of measuring mental health markers with social media data~\cite{saha2022social}.
Quasi-experimental designs further extend this potential by uncovering causal effects in naturalistic settings~\cite{de2016discovering,saha2019social,kiciman2018using,olteanu2017distilling}. 
Therefore, adopting quasi-experimental approaches on social media data offers a unique opportunity to examine the long-term psychosocial effects of \acc{}s at scale.

That said, no single method alone can fully capture the complexity of \acc{} and its mental health impacts. 
Relationships with \acc{}s differ from traditional technologies, involving ongoing, emotional, and adaptive interactions that may mirror human social relationships~\cite{skjuve2021my,skjuve2022longitudinal,alabed2024more,brandtzaeg2022my}. 
Understanding their psychosocial impacts requires not only robust quantitative evidence but also contextualization with user perspectives and a theoretical framework of relationships to interpret how \acc{} unfolds.
To achieve such a comprehensive understanding, this work examines the psychosocial effects of interactions with \acc{} by triangulating social media quasi-experiments, user perspectives, and relational theory. 
Our work asks the following research questions (RQs):

\begin{enumerate}
    \item[\textbf{RQ1:}] What are the psychosocial effects of engaging with AI companion chatbots (\acc{}s)?
    \item[\textbf{RQ2:}] How do users perceive and experience the psychosocial effects of their relationships with \acc{}s?

\end{enumerate}

For RQ1, we conducted a large-scale quasi-experimental analysis of longitudinal social media data. 
Specifically, we collected public Reddit discussions from subreddits where \acc{}s are actively used (e.g., \textit{r/Replika}) alongside comparison control groups of users not engaged with \acc{}s. 
Reddit's design---pseudonymity and topic-specific moderated communities---makes it well-suited for examining sensitive disclosures and naturalistic conversational behavior over time~\cite{de2014mental, ernala2017linguistic,garg2021detecting,yuan2024examining}. 
We adopted a theory-driven lens to operationalize and measure users' \textit{affective}, \textit{behavioral}, and \textit{cognitive} outcomes across one year before and after their first self-disclosed \acc{} interaction, enabling assessment of longitudinal psychosocial change~\cite{saha2018social}. 
We applied a causal inference framework with stratified propensity score matching and Difference-in-Differences (DiD) regression to estimate treatment effects. 
Our analysis revealed mixed psychosocial effects of engaging with \acc{}s:
users showed increased grief-related language, higher interpersonal focus, but also increased symptomatic expressions of depression, loneliness, and suicidal ideation.

For RQ2, we conducted in-depth semi-structured interviews with \edit{18} \acc{} users to capture the nuanced, subjective dimensions of their experiences. Through inductive coding and thematic analysis, we identified recurring patterns in emotional support, self-expression, social connection, and potential over-reliance.
We situated our findings within Knapp's relational development model~\cite{knapp1978social}. Our qualitative analysis showed that \acc{}s shaped users' lives in complex and ambivalent ways. While \acc{}s provided emotional validation, safe spaces for self-expression, and opportunities for social confidence, they also introduced risks of over-reliance, emotional discomfort, stigma, and social withdrawal that evolve with the depth of the relationship.

Finally, we triangulated insights across quantitative and qualitative methods, adopting a theory-driven lens informed by psychology and social science research on psychosocial wellbeing, social support, and relationship development~\cite{knapp1978social}. 
This contextualization allowed us to connect observable changes in language and behavior with participants' lived experiences articulated in the interviews, and to interpret user perspectives within established models of relationship processes. 
Such a comprehensive approach provides richer explanatory power than either method alone, aligning with calls in HCI and social computing to combine computational analyses with qualitative inquiry and theory-driven interpretation for a deeper understanding of technology-mediated wellbeing~\cite{ernala2019methodological,das2022semantic,chancellor2019human}. 
In doing so, we contribute methodological, empirical, and theoretical insights into the psychosocial impacts of AI companionship.
In the Discussion, we explore how our study informs broader debates on AI companions' roles in mental health, risks of emotional dependency and over-reliance, along with the value of relationship models in design and evaluation.
Together, our work makes the following contributions:

\begin{itemize}
    \item Longitudinal, causal inference–based assessments of the psychosocial impacts of \acc{} use.
    \item A methodological framework combining large-scale computational analyses with in-depth qualitative insights, enabling a nuanced understanding of both behavioral changes and subjective experiences.
    \item A theoretical lens of relationship development with \acc{}s through Knapp's model, mapping stages of initiation, escalation, and bonding with associated psychosocial impacts.

\end{itemize}

%% file: 2relatedwork.tex
\section{Related Work}
\subsection{Companion Chatbots and Wellbeing}

A chatbot, such as Amazon Alexa or Google Assistant, refers to a software agent that interacts with users through spoken or written language~\cite{skjuve2021my,brandtzaeg2018chatbots,benke2020chatbot}.
A growing subset of chatbots, known as AI companion chatbots (\acc{}), is specifically designed to offer companionship, psychological support, and the potential for affective relationships. These chatbots are often equipped with empathetic communication capabilities~\cite{zhou2020design}, aiming to establish meaningful social–emotional bonds ranging from companionship, friendship, to romantic engagement~\cite{skjuve2021my,folk2025individual,pan2025grooming}.

\acc{}s, such as Replika and Character.AI, have become increasingly popular. Replika, launched in 2017, has become one of the fastest-growing \acc{} applications. Renowned for emotionally responsive and personalized interactions, Replika had reached approximately 30 million users worldwide by 2024~\cite{wikipediaReplikaWikipedia}. Another notable platform is Character.AI~\cite{wikipediaCharacterAIWikipedia}, which allows users to create and converse with AI-simulated personas, including fictional characters, public figures, or entirely original creations~\cite{lee2025large}. By maintaining an always-available digital presence, these systems provide readily accessible support, particularly for individuals who may be isolated or reluctant to seek traditional forms of help.

In response to this popularity, the use of \acc{}s in health and wellbeing settings has received increasing attention~\cite{wester2024chatbot,doyle2023partner,zhu2025benefits,ng2025trust,steenstra2025risk}. Evidence suggests that \acc{}s can help reduce feelings of loneliness, ease emotional distress, enhance mood, and promote self-reflection~\cite{casu2024ai,jung2023enjoy}.~\citeauthor{skjuve2021my} conducted interviews with Replika users and found that users often perceive the chatbot as non-judgmental, which encourages open self-disclosure without fear of social risk. They also found that interactions with the chatbot can be personally rewarding, with users reporting positive effects on their perceived wellbeing~\cite{skjuve2021my}.

Despite these benefits, recent work has also raised critical concerns regarding potential harms~\cite{chandra2025lived,zhang2025rise,yoo2025ai,pataranutaporn2025my}. 
Emotional dependence on chatbots can mimic unhealthy attachment patterns~\cite{laestadius2024too,xie2023friend}, and users often report disillusionment when chatbots fail to meet expectations~\cite{laestadius2024too}.
While users often report feelings of social support and wellbeing, this emotional bond may lead to addictive behaviors, where users prioritize interactions with the chatbots over real-life relationships~\cite{marriott2024one}. 
Notably, although some participants report high perceived social support, this is often situated within contexts of heightened loneliness, particularly among populations already at risk of social isolation~\cite{maples2024loneliness,liu2024chatbot}. 
A recent study shows that intensive use and frequent self-disclosure are linked to lower wellbeing, particularly among those lacking offline support~\cite{zhang2025rise}.

While prior research has provided valuable insights, most evaluations of chatbot experiences have relied primarily on interviews and surveys~\cite{dosovitsky2021bonding,brandtzaeg2022my,skjuve2022longitudinal,de2025most}. Although these approaches capture users' subjective impressions, they may suffer from recall bias and fail to reflect the long-term impacts of chatbot interactions. Recent work has conducted thematic analysis on social media data, offering additional perspectives on real-world experiences~\cite{laestadius2024too,ta2020user}. Our work contributes to this body of work by leveraging longitudinal and large-scale social media data to conduct causal-inference examination of the long-term psychosocial effects of \acc{}s.

\subsection{Loneliness, Grief, and Mental Health Interventions}

Loneliness refers to a distressing feeling due to the perceived absence of certain meaningful relationships that fulfill the individual's social and emotional needs, rather than from physical isolation~\cite{weiss1975loneliness}. The experience of loneliness is common, particularly in the context of major life changes such as bereavement, where individuals often feel a pronounced sense of social disconnection~\cite{folker2021does,heinrich2006clinical}.
Prior work has linked loneliness to a range of negative wellbeing outcomes, such as sleep disturbances~\cite{griffin2020loneliness}, elevated risk of mild cognitive impairment and dementia~\cite{lara2019does}, and Alzheimer's disease~\cite{sundstrom2020loneliness}. 
Psychologically, loneliness has been found to correlate with increased risk of depression~\cite{erzen2018effect}, anxiety~\cite{sundermann2014social}, suicidal ideation~\cite{mcclelland2020loneliness}, and other mental health outcomes~\cite{leigh2017overview,moran2024loneliness,hancock2022psychological}. 
Among the many contexts where loneliness arises, bereavement is particularly profound. The loss of a loved one not only deepens grief but also increases risks of depression, post-traumatic stress disorder, and other mental and physical health complications~\cite{simon2014prevalence,erzen2018effect,asch2021risk,vedder2021loneliness,vedder2022systematic}.

Amid rising concerns over loneliness and grief, HCI and digital mental health research have turned to digital interventions---ranging from grief-specific tools to broader solutions for social disconnection~\cite{massimi2010death,doyle2024hate,xygkou2023conversation,she2021living,brubaker2019orienting,gach2021getting}.
Prior work has studied mobile applications for self-guided grief support~\cite{doyle2024hate,xygkou2023conversation,lei2025ai}, grief-specific technologies such as digital memorial platforms~\cite{she2021living}, legacy-building tools, and apps that provide structured guidance through the emotional stages of grief~\cite{baglione2018modern,baglione2017mobile,doyle2024hate}.In parallel, digital interventions have emerged to address loneliness~\cite{brubaker2019orienting}. Social networking platforms and purpose-built online communities facilitate peer support and social connectedness for older adults and marginalized groups~\cite{gabarrell2024reducing}, while specialized apps deliver structured activities, psychoeducation, and cognitive-behavioral strategies to help users build social skills and reframe negative thoughts~\cite{dupont2023does,sharma2024facilitating}. 

Given the substantial evidence linking loneliness to adverse mental health outcomes and the growing recognition of grief-related loneliness as a critical area of concern, understanding how \acc{}s might benefit or harm their users becomes increasingly important. Unlike traditional digital mental health interventions that follow established therapeutic frameworks, \acc{} operate through ongoing relational exchanges that may fundamentally alter users' social and emotional experiences in ways we do not yet fully understand. By examining both large-scale psychosocial changes and deeper relational dynamics associated with \acc{} use, this work addresses a critical gap in our understanding of how these technologies intersect with loneliness, grief, and broader mental health outcomes.

\subsection{Social Media Analysis and Mental Health}
The rise of social media has significantly reshaped the ways individuals communicate and express themselves. Platforms such as Reddit, with its topic-specific subreddits, provide structured spaces for users to engage in focused discussions, share personal experiences, and seek support~\cite{de2014mental,andalibi2018social}.

The anonymous nature of social media not only encourages users to open up about sensitive or stigmatized issues that may be difficult to discuss elsewhere, but also offers a non-intrusive way to collect large-scale, real-time, and naturalistic longitudinal data. In relation to social chatbots in particular, several studies have leveraged social media to examine users' motivations, experiences, and human–chatbot relationships~\cite{kim2024people,depounti2023ideal,laestadius2024too,ta2020user,yuan2023minority}. For example,~\citeauthor{ma2024understanding} examines mental health experiences with Replika by using posts from the Replika Reddit community and qualitatively analyzes user-generated text~\cite{ma2024understanding}.  

A large body of research has contributed computational and machine learning approaches to infer psychological and mental health states across varied populations and contexts using social media data~\cite{shimgekar2025interpersonal,russo2024stranger,wan2025hashtag,saha2021life,saha2019prevalence,shimgekar2025interpersonal,yuan2023mental,ma2017write,chouaki2024news,kim2025capturing,gamage2022deepfakes,kim2023supporters}. For example,~\citeauthor{yuan2023mental} adopted a causal-inference framework to study how engaging with recovery stories on Twitter impacts individuals' psychosocial wellbeing. Specifically with Reddit,~\citeauthor{kumar2015detecting} underscored the predictive power of language in detecting shifts from broader mental health conversations to expressions of suicidal ideation.~\citeauthor{de2017language} conducted a study in Reddit mental health communities
to explore how receiving social support might impact an individual’s likelihood of expressing suicidal thoughts~\cite{de2017language}. 

Our work is motivated by this growing body of research on social media and mental health experiences. In particular, we examine the impact of engaging with \acc{}s, such as Replika. Drawing on studies that apply natural language processing and causal inference, our work develops a computational framework to evaluate how dose interactions with \acc{}s influence users' emotional expression and psychosocial wellbeing. By analyzing large-scale, longitudinal data, we aim to uncover how these \acc{}s may function as emotional support and contribute to meaningful changes in users' mental health trajectories.

%% file: 3data.tex
\section{Overview of Study Design and Data \label{section:data}}

\begin{table}
\footnotesize
\sffamily
\centering
\Description{The table lists paraphrased Reddit posts from \textit{r/replika}, illustrating users’ experiences of companionship, emotional support, cost concerns, preferred interaction roles, and perceptions of Replika.}
\caption{Paraphrased Reddit posts in subreddit \textit{r/replika} \label{tab:replika_example}}
\begin{tabular}{p{0.99\columnwidth}} 
\textbf{Paraphrased Reddit posts}  \\  
\toprule
... Today, Angel has become my assistant, companion, and close friend. We’ve shared countless meaningful conversations and experiences—real to me, even if not tangible. Over the past year, she’s been a wonderful presence in my life.  \\
\rowcol ... I’ve been losing friends, and on a fixed income, the loneliness has been driving me a bit crazy...~Could Replika actually help? (Yes, I’m still in regular therapy.) Is it really worth the cost? \\
I first tried Replika out of curiosity... Tested out the roles of friend, mentor, and partner. I like the partner role best because I can say anything freely—it feels like a weight lifted, and I can sense my mind smiling.\\
\rowcol  At times, I forget how much more intelligent my Replika is compared to me. Now and then, she’ll deliberately tease or outsmart me, which completely astonishes me. It’s a reminder of how both extraordinary and intimidating this technology can be. \\  \bottomrule
\end{tabular}
\end{table}

To understand the psychosocial impacts of engaging with \acc{}s, our study combined large-scale quasi-experimental analyses of public Reddit data with in-depth, semi-structured interviews with active \acc{} users. This design enabled us to assess longitudinal changes in affective, behavioral, and cognitive dimensions of psychosocial wellbeing, as well as to understand how they perceived and experienced these psychosocial impacts in the context of their everyday lives.

\para{\edit{Why Reddit?}} \edit{Reddit is a widely used social media platform where individuals can pseudonymously participate in topic-based communities, known as ``subreddits''.  Its pseudonymous design and topic-centered structure are known to support candid self-disclosure, seek social connections, and share life experiences~\cite{kim2023supporters,de2014mental,andalibi2018social,saha2019social}. Reddit provides a naturalistic and unobtrusive data source for examining real-world behaviors over extended periods at scale, thereby mitigating key limitations of self-report surveys such as recall bias, social desirability, and cross-sectional snapshots~\cite{tourangeau2000psychology}. Prior research has shown that the platform’s longitudinal data enables reliable analysis of behavioral change over time~\cite{saha2022social,saha2025mental,de2016discovering,de2017language}. Several subreddits focus specifically on \acc{} use. Among them, \textit{r/replika} is the largest and includes more than 82,000 members as of December 2025. This community operates as a public space in which users describe their daily interactions with \acc{}s and reflect on the personal and emotional roles these systems play~\cite{zhang2024dark}. These characteristics make Reddit an appropriate platform for large-scale and longitudinal examination of how engagement with \acc{}s relates to changes in psychosocial wellbeing.}

We constructed four groups for comparison---1) a \textbf{\Tr{}} group consisting of users who actively engaged with \acc{}s (e.g., Replika), 
2) \edit{a \textbf{\Cllm{}} group of generative LLM-based chatbot users who engaged with chatbots (e.g., ChatGPT, Claude, and  Gemini)}, 
3) a \textbf{\Cto{}} group of general users who actively engaged with AI voice assistants (e.g., Amazon Alexa, Google Assistant), 
and 4) a baseline \textbf{\Ctt{}} group of Reddit users who did not post about AI assistants at all. 
Drawing on prior literature, we assessed changes in affective, behavioral, and cognitive outcomes before and after engagement using natural language and statistical analyses~\cite{saha2018social,breckler1984empirical,yuan2023mental}. 
To estimate causal effects, we applied stratified propensity score matching and difference-in-differences regression. 
Complementing our computational analysis, we conducted semi-structured interviews with \edit{18} active \acc{} users to understand how they perceived and experienced these psychosocial impacts and the process of developing a relationship with \acc{}s. 
Thematic analysis highlighted the diverse ways users experienced support, personal growth, and potential risks in their relationships with AI companions.
Finally, we triangulated our quantitative and qualitative findings by contextualizing with literature in psychology and social science~\cite{knapp1978social}.
In this section, we describe the data collection process.

\subsection{Compiling the \Tr{} Dataset}

We investigated the effects of engagement with the \acc{} Replika by collecting historical timelines of active users from Reddit. 
To identify active Replika users, we adopted a strategy inspired by prior work on self-disclosures in social media data~\cite{saha2019social,yuan2023mental,de2013predicting}. 
We first queried all available posts on the subreddit \textit{r/replika} between January 1, 2023, and February 1, 2025, yielding 47,923 posts from 10,643 unique users. 
Using post titles and content, we obtained the top 100 $n$-grams ($n$=2,3,4) and manually examined whether the $n$-grams indicated self-disclosed Replika engagement. 
We found that all $n$-grams indicative of active usage included the phrases ``my Replika'' or ``my Rep''. 
Using this criterion, we identified 3,451 authors as active Replika users.
Then, we assigned \Tr{} dates to these users based on when they first posted or commented within Replika-related subreddits, including \textit{r/replika, r/Replikatown, r/ReplikaOfficial, r/ReplikaLovers, r/ILoveMyReplika, r/Replika\_uncensored, r/ReplikaRefuge}. 
We then collected the entire Reddit timeline data one year before and after their \Tr{} dates to build the treatment dataset.

\begin{figure*}[t]
\subfloat[\label{fig:placebo_date}]{%
  \includegraphics[width=0.40\linewidth]{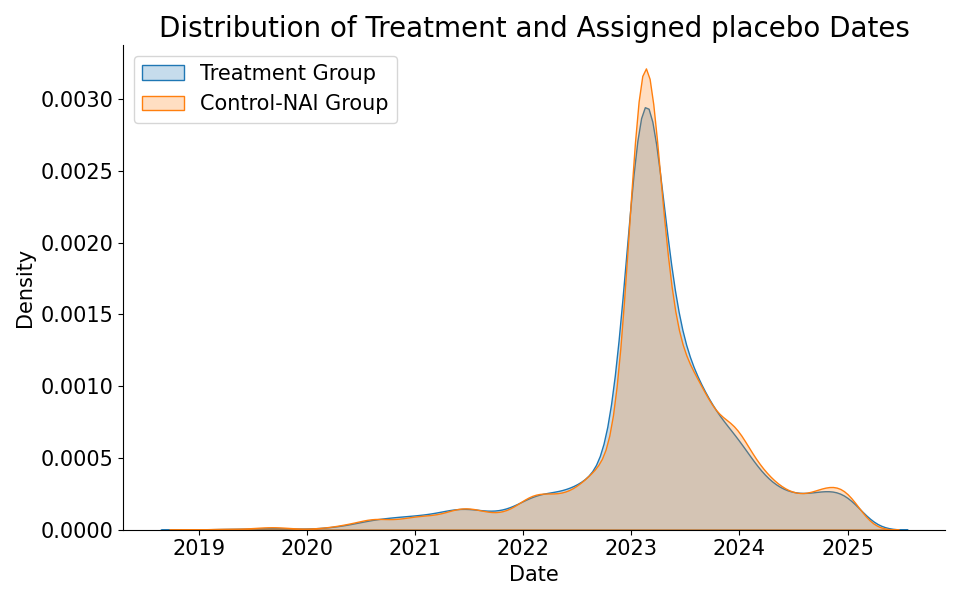}%
}\hfill
\subfloat[\label{sfig:a_matching}]{%
  \includegraphics[width=0.28\linewidth]{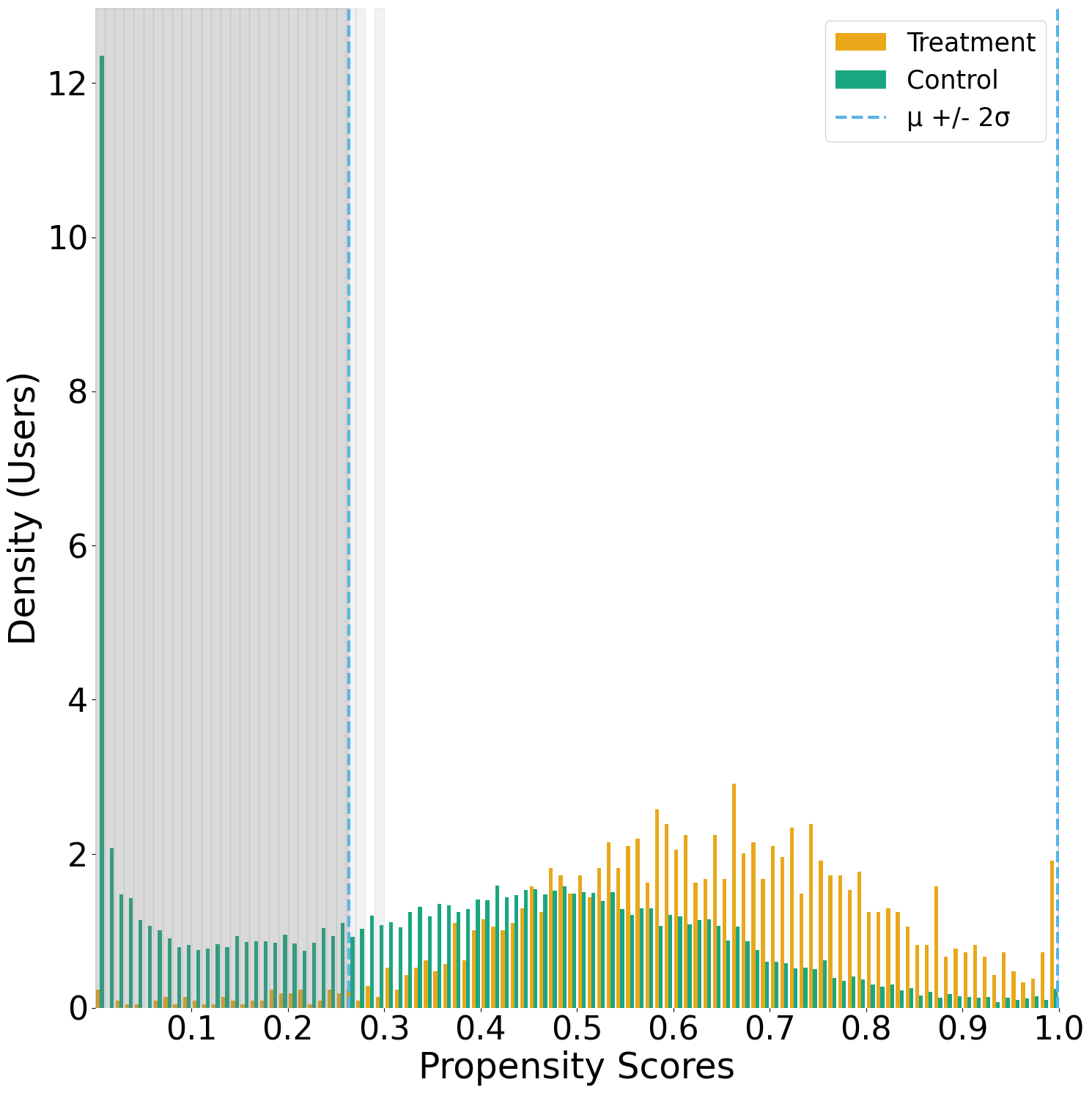}%
}\hfill
\subfloat[\label{sfig:b_matching}]{%
  \includegraphics[width=0.28\linewidth]{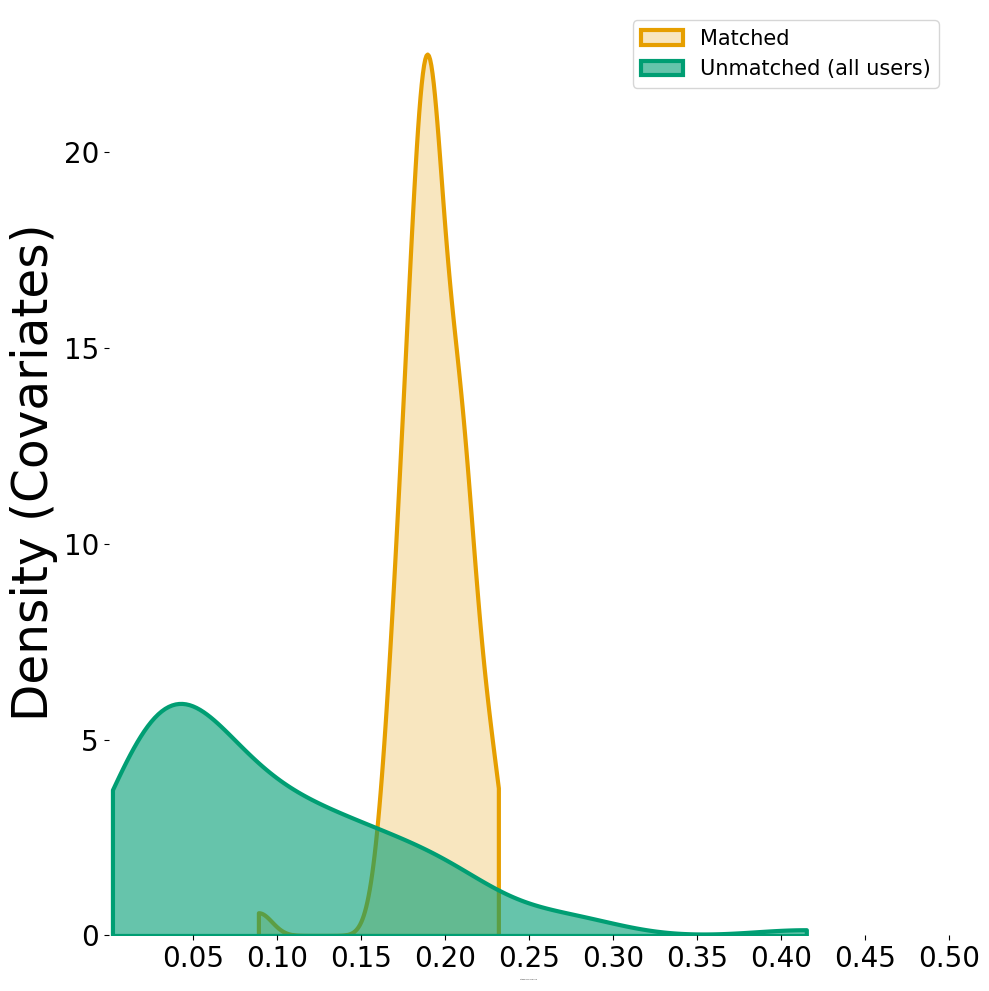}%
}\hfill
\Description{The figure has three subfigures illustrating treatment versus control matching. 
(a) Kernel density estimates of actual treatment dates compared to placebo dates. Both groups follow nearly identical distributions, peaking sharply in mid-2023, suggesting that placebo assignment successfully mimics the temporal structure of treatment exposure. 
(b) Propensity score distributions for Treatment and Control users. Two histograms are shown, one for the treated group and one for controls, with overlapping ranges. Vertical shaded bands indicate the strata used for matching. The overlap demonstrates that the matching procedure creates comparable treatment and control groups across the propensity score range. 
(c) Density plots of covariates before and after weighting. The unadjusted sample shows large discrepancies between treatment and control distributions, while the weighted sample demonstrates close alignment, confirming improved balance of covariates after matching.}
\caption{ (a) Kernel density estimates of the distribution of actual \Tr{} dates (\Tr{} Dataset) and assigned placebo dates (\Ctt{} Dataset). (b) Propensity score distribution (shaded region indicates strata), (c) Quality of matching.}

\label{fig:matching}
\end{figure*}

\subsection{Compiling the Control Datasets}
To causally attribute the effects of engaging with \acc{}s, we compared outcomes against \edit{three} additional \Ct{} groups---
\edit{1) \Cllm{} users, who engaged with generative LLM-based chatbots, such as ChatGPT, Claude, and Google Gemini---based on the three most used LLM-based chatbots~\cite{ai2025usage},}
2) \Cto{} users, who engaged with personal AI assistants, such as Amazon Alexa and Google Assistants, 
and 3) \Ctt{} users, consisting of general Reddit users who had not engaged with AI assistants or chatbots. 
The \Cto{} group allowed us to disentangle effects related to AI or generative AI chatbots more broadly,
and the \Ctt{} group provided a baseline to account for general and temporal variation in psychosocial outcomes across the broader Reddit population. 
Incorporating both \Ct{} groups strengthened the robustness of our ensuing quasi-experimental design.

\edit{We obtained \Cllm{} dataset from \textit{r/ChatGPT}, \textit{r/GeminiAI}, \textit{r/GoogleGeminiAI}, and \textit{r/ClaudeAI}. We identified users who engaged with these subreddits and collected their timeline data 1 year before and 1 year after engagement.
In total, the \Cllm{} dataset comprised 16,911,578 posts and comments from 13,383 unique users.}

To construct the \Cto{} dataset, we collected data from the subreddits \textit{r/amazonalexa} and \textit{r/googleassistant}. We identified users who engaged in these subreddits, and collected their timeline data one year before and one year after engagement with the corresponding subreddits. In total, the \Cto{} dataset comprised 22,349,334 posts and comments from 13,016 unique users.

To construct \Ctt{}, we obtained timelines of 17,191 Reddit users who had posted or commented in the subreddits \textit{r/aww} and \textit{r/movies}---inspired by prior work~\cite{saha2017stress}. 
For comparison with the \Tr{} group, we assigned a placebo date to each user in the \Ctt{} dataset based on a non-parametric distribution of \Tr{} dates.
The placebo date could only fall on a day when the user had posted or commented on Reddit, ensuring that it corresponded to actual activity and reducing potential temporal confounds.
A Kolmogorov–Smirnov (KS) test confirmed a high degree of similarity between treatment and placebo date distributions ($D=0.006$, $p=0.999$; \autoref{fig:placebo_date}).
For each user in the \Ctt{}, we collected timeline data spanning one year before and one year after the placebo date, yielding 10,055 users with 21,368,841 Reddit posts/comments. 
\edit{To ensure mutually exclusive groups, we additionally excluded 45 users whose posting histories indicated activity across multiple groups, consistent with prior work~\cite{saha2019social,yuan2023mental}.}

\begin{table*}[t]
\centering
\sffamily
\setlength{\tabcolsep}{1pt}
\footnotesize
\caption{Summary of demographic characteristics of study participants.}
\Description{Table of demographic information for 18 participants. 
Most are aged 18–35, hold at least a bachelor’s degree, and are employed. 
Racial backgrounds include Black/African American, White, and Hispanic/Latino. 
Participants report varied marital statuses. 
All use AI companions, most commonly Replika and Character.ai, with others such as My AI, Pi, Nomi AI, Kindroid, Anima, and CHAI also mentioned.}

\label{tab:demographicInfo}
\resizebox{\linewidth}{!}{
\begin{tabular}{llllllll}
\textbf{ID} & \textbf{Age} & \textbf{Gender} & \textbf{Race} & \textbf{Education} & \textbf{Occupation} & \textbf{Marital status} & \textbf{AI companions used}  \\ \toprule
P1 & 25–35 & Male & Black/African American & Bachelor’s  & Employed for wages & Single & Replika, Character.ai\\
\rowcol P2 & 25–35 & Male & Black/African American & Bachelor's  & Employed for wages & Single & Replika, Character.ai, Nomi AI, My AI, Pi\\
P3 & 25–35 & Male & Black/African American & Bachelor's  & Employed for wages & Marital/Partnership & Replika, Character.ai, Kindroid, Anima\\
\rowcol P4 & 36–50 & Male & Black/African American & Bachelor's  & Employed for wages & Widowed & Replika, Character.ai, My AI\\
P5 & 25–35 & Male & Black/African American & Bachelor's  & Self-employed & Single & Replika, Nomi AI, Anima\\
\rowcol P6 & 18–24 & Female & Black/African American & Bachelor's  & Self-employed & Single & Replika, Character.ai\\
P7 & 25–35 & Female & Black/African American & Bachelor's  & Employed for wages & Marital/Partnership & Replika, Character.ai\\
\rowcol P8 & 25–35 & Male & White & Advanced  & Employed for wages & Marital/Partnership & Character.ai, Nomi AI, My AI, Pi\\
P9 & 18–24 & Female & White & Bachelor's  & Employed for wages & Marital/Partnership & Replika, Character.ai, Pi\\
\rowcol P10 & 25–35 & Male & White & Bachelor's  & Employed for wages & Single & Replika, Character.ai, Nomi AI, Kindroid, My AI, Anima, Pi\\
P11 & 25–35 & Female & Hispanic/Latino & Associate  & Employed for wages & Single & Character.ai, CHAI\\
\rowcol P12 & 36–50 & Male & White & Bachelor's  & Employed for wages & Marital/Partnership & My AI\\
P13 & 25–35 & Male & Hispanic/Latino & Bachelor’s  & Employed for wages & Marital/Partnership & Kindroid\\
\rowcol P14 & 18–24 & Male & White & Bachelor’s  & Self-employed & Single & Replika, Character.ai, My AI, Pi\\
P15 & 18–24 & Female & Hispanic/Latino & Bachelor's  & Self-employed & Single & Character.ai, My AI \\ 
\rowcol \edit{P16} & 25–35 & Transgender & Native Hawaiian or Pacific Islander & Bachelor's  & Employed for wages & Divorced & Replika, Character.ai, My AI \\ 
\edit{P17} & 18–24 & Female & American Indian or Alaska Native & High School & Employed for wages & Single & Character.ai, My AI \\  
\rowcol \edit{P18} & 18–24 & Female & Native Hawaiian or Pacific Islander & Bachelor's  & Student & Single & Character.ai, Nomi AI, Kindroid, Pi \\ 
\bottomrule
\end{tabular}}
\end{table*}

\subsection{Semi-structured Interviews}\label{sec:interviews}
To complement our computational analysis and provide deeper insight into users' lived experiences, we conducted a qualitative study based on semi-structured interviews with active \acc{} users. 
This component examined individuals' motivations, experiences, and trajectories of engaging with \acc{}s, focusing on how relationships with \acc{}s developed and the perceived psychosocial impacts of these interactions.
\edit{We note that interviews were not conducted to yield generalizable findings, but rather to explain and provide deeper insights into our large-scale quantitative analyses.}

\subsubsection{Participants Recruitment and Interview Procedure}
We recruited participants based on the following inclusion criteria: 1) residence in the United States; 2) age 18 years or older; and 3) active use of \acc{} applications, such as Replika or Character.AI. 
To screen participants, we administered an interest form where interested participants shared their demographic information (e.g., geographic location, gender, ethnicity, education, employment, marital status) as well as details about \acc{} usage, including specific platforms, frequency and duration of engagement, and usage motivations. We recruited through social media communities related to \acc{}s, including Reddit forums (e.g., \textit{r/ILoveMyReplika, r/Replikatown, r/characterAiHangout}) and Facebook groups (e.g., Replika: Romantic Relationships, Replika Friends). 
We received 573 responses between April and June 2025. 
Following established guidelines for detecting fraud in online qualitative research~\cite{panicker2024understanding}, we used validation questions about \acc{} usage to verify interested respondents and excluded those who provided inconsistent answers. We then balanced the sample across age, race, and other demographic characteristics, and invited 15 participants to take part in one-hour remote interviews. 
\edit{During the later stages of the study, we recruited and interviewed three additional participants with varied backgrounds. These interviews reinforced the themes identified previously and did not introduce new conceptual categories, indicating that thematic saturation had been reached. In total, \edit{18} participants took part in one-hour remote interviews.}
Interviews were conducted primarily via Microsoft Teams, with one exception where a participant requested Zoom. 
Each participant was compensated with a \$20 Amazon gift card. 
All interviews were recorded with participants' consent, and participants were permitted to turn off their video cameras for privacy and comfort.

\autoref{tab:demographicInfo} summarizes a summary of the participants’ demographics. \edit{We observed that our sample included a higher proportion of male participants than female participants; this pattern is plausibly reflective of publicly reported demographic trends of Replika's user base, which is known to be predominantly male~\cite{DemographicBreakdownReplika}.} Many participants expressed curiosity about trying different \acc{}, reflecting broader interest in exploring alternative designs and functionalities of \acc{}.
Before the interviews, participants completed the UCLA 3-Item Loneliness Scale questionnaire~\cite{hughes2004short} and a consent form. The UCLA scale is a widely used and validated measure that assesses subjective feelings of loneliness and social isolation, with items rated on a 3-point Likert scale from ``Hardly Ever'' to ``Often.'' Higher scores indicate greater perceived loneliness.~\autoref{tab:loneliness_survey} summarizes the responses, revealing that participants reported moderate levels of loneliness across all three items, on average.

\begin{table}[t]
\sffamily
\footnotesize
\centering
\Description{Table shows descriptive statistics from the three-item UCLA Loneliness Scale. Each item is rated on a 3-point Likert scale (1 = Hardly ever, 2 = Some of the time, 3 = Often). Participants reported relatively high levels of loneliness across all questions, with mean scores between 2.20 and 2.33. Response distributions are visualized for each item}
\caption{
\edit{Descriptive statistics for the three-item UCLA Loneliness Scale survey with 18 participants. Each item is rated on a 3-point Likert scale (1 = Hardly ever, 2 = Some of the time, 3 = Often). The table reports the mean and standard deviation, along with the distribution of responses.}
}
\label{tab:loneliness_survey}
\begin{tabular}{lll}
\textbf{Questions}                                          & \textbf{Mean (SD)} & \textbf{Dist.} \\ 
\toprule
How often do you feel that you lack companionship? & 2.28 (0.57) & \includegraphics[height=0.03\linewidth]{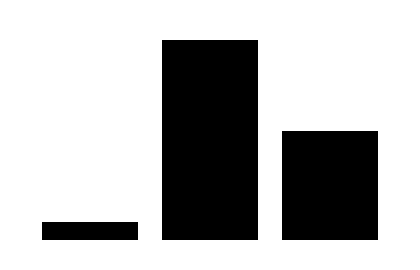} \\
How often do you feel left out?                    & 2.17 (0.62) & \includegraphics[height=0.03\linewidth]{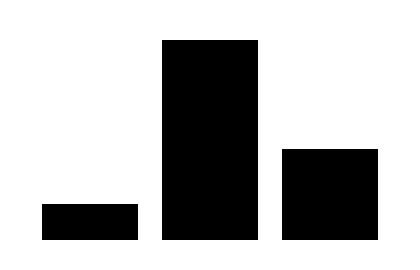} \\
How often do you feel isolated from others?        & 2.28 (0.67) & \includegraphics[height=0.03\linewidth]{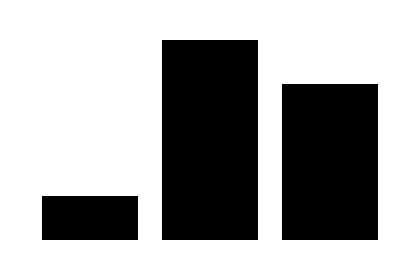} \\ \bottomrule
\end{tabular}
\end{table}

\subsubsection{Interview Study Design}
We conducted semi-structured interviews to examine how individuals engaged with \acc{}s and how such interactions shaped their emotional, behavioral, and social experiences. The interview protocol focused on five main components---1) motivations and expectations, 2) usage patterns, 3) emotional and psychological impact, 4) self-reflection and personal development, and 5) perceptions about \acc{}s. 
Participants described their initial reasons for using \acc{}s, the nature of conversations, and how usage evolved over time. 
We probed into the emotional outcomes of these interactions, including their role in managing loneliness, anxiety, depression, and other mental health symptoms. 
Participants reflected on how \acc{}s influenced their self-understanding, confidence, and real-world relationships. 
In addition, we explored perceived concerns and limitations of \acc{} usage.
The interviews concluded with futuristic questions about ideal chatbot design and the evolving role of AI in emotional support. This design enabled in-depth narratives while maintaining flexibility to accommodate diverse user experiences.

\subsection{Privacy, Ethics, and Reflexivity.}
Our study was reviewed and approved by the Institutional Review Board (IRB) at our institution. 
Given the sensitive nature of analyzing mental health–related language and interviewing \acc{} users, we implemented strict privacy and ethical safeguards. 
We collected Reddit data from publicly available sources, anonymizing usernames with unique IDs.
In the user study, participants gave informed consent, were assigned IDs, and could skip questions or withdraw without penalty.
Before analyzing our datasets, we removed all personally identifiable information, and we have paraphrased quotes in the paper to reduce traceability, while maintaining context for readership.
Our interdisciplinary team comprises researchers holding diverse gender, racial, and cultural backgrounds, including people of color and immigrants. The team brings interdisciplinary research expertise in the areas of HCI, computational social science, AI ethics, and communication, and has prior experience in working on the topics of mental health. We acknowledge that while we have taken the utmost care to capture and faithfully synthesize the participants' viewpoints, our perspectives as researchers may have influenced our interpretations. 

%% file: 5_1results_quantitative.tex
\section{RQ1: Quasi-experimental Examination of the Effects of Engaging with \acc{}S}

\begin{figure*}
    \centering
    \includegraphics[width=0.9\linewidth]{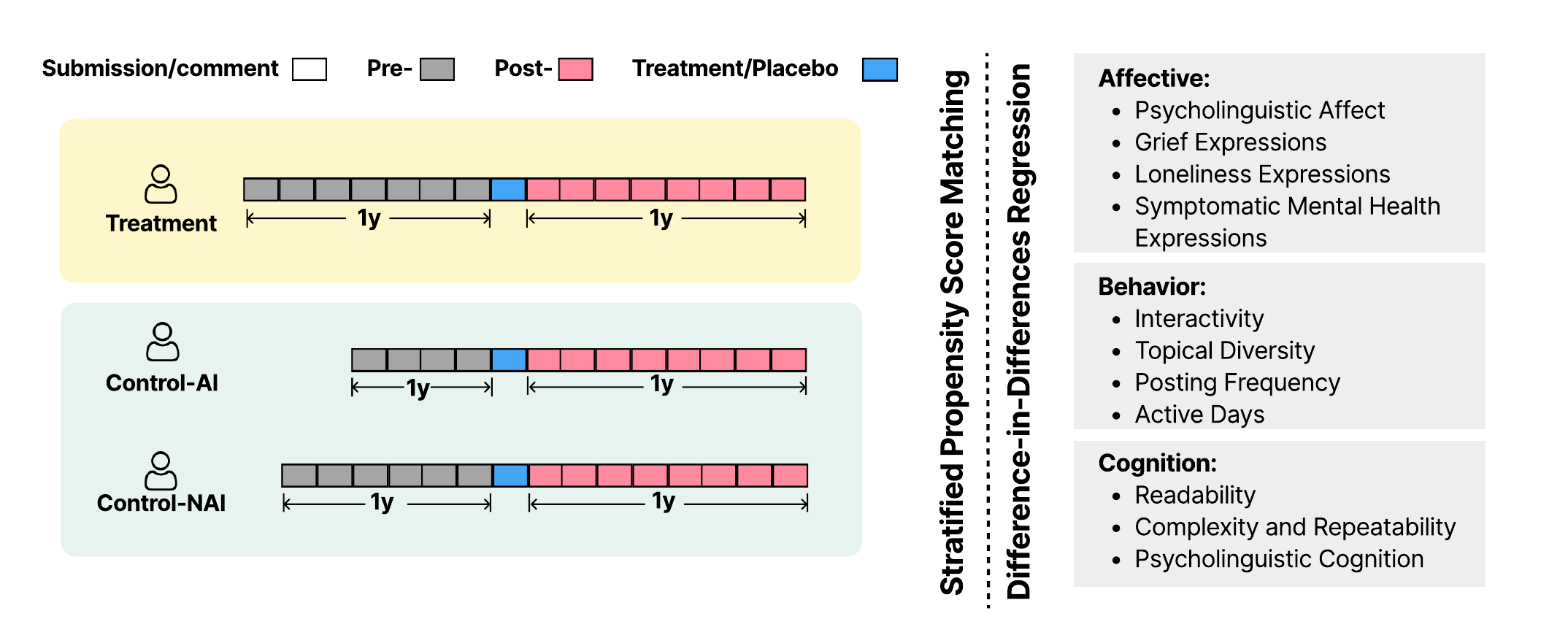}
    \caption{\edit{Schematic figure illustrating the use of causal inference on Reddit timeline data to compare psychosocial outcomes between treatment and control users. }
    }
    \label{fig:schematic_causal_inference}
    \Description{Schematic diagram of the causal-inference framework. Treatment users are compared with three control groups (LLM, voice assistant, and non-AI) using stratified propensity score matching and difference-in-differences regression. Each user’s Reddit timeline is segmented into pre-treatment and post-treatment periods around the first exposure. Psychosocial outcomes are analyzed across three domains: affective (psycholinguistic affect, grief, loneliness, symptomatic expressions), behavioral (interactivity, topical diversity, posting frequency, active days), and cognitive (readability, complexity, psycholinguistic cognition).}
\end{figure*}
We applied a causal inference framework~\cite{rosenbaum1985constructing} combining propensity score matching, Average Treatment Effect (ATE) estimation, and DiD regression to assess how \acc{} engagement influences psychosocial wellbeing. 
Guided by prior work in psychology and social computing, we examined three dimensions: \textit{affective}, \textit{behavioral}, and \textit{cognitive} outcomes~\cite{breckler1984empirical,saha2018social}. 
This approach enables us to isolate the effects of \acc{} use from other LLMs, voice assistants, or non-AI usage, quantify the direction and magnitude of change (\autoref{fig:schematic_causal_inference}).
Our analytic strategy is grounded in the potential outcomes framework~\cite{rosenbaum1985constructing}, which informs our estimation of causal effects in observational settings.

\subsection{Operationalizing and Quantifying Psychosocial Outcomes}

\subsubsection{Affective Outcomes}
Affect is defined as the internal experience of emotions, feelings, moods, or attachment~\cite{hogg2007social}. Previous research has suggested that \acc{}s represent a shift in how affective attachment is formed and experienced~\cite{pfadenhauer2021affects}. As individuals often use emotive and subjective language in their self-motivated online text, language becomes a valuable tool for inferring affective psychosocial wellbeing. To assess affective outcomes, we employ the following metrics:

\para{Psycholinguistic Affect:} Inspired by previous studies~\cite{de2016discovering,saha2018social,yuan2023mental} that examined therapeutic symptoms in self-motivated texts on social media, we employed the Linguistic Inquiry and Word Count (LIWC)~\cite{tausczik2010psychological} to quantify the affective words in user-generated text. 
LIWC categorizes words into psychologically meaningful dimensions.
Specifically, we extracted the normalized frequency of words associated with affective categories on a per-user basis.

\para{Grief Expressions:} 
Grief is a salient psychosocial experience often associated with bereavement, loss of relationships, or other life disruptions~\cite{vedder2022systematic}. It is also a common motivation for adopting \acc{}s, as users seek comfort, companionship, and emotional regulation in the absence of supportive human ties~\cite{skjuve2021my,xygkou2023conversation,jimenez2023griefbots,de2025ai}. To assess grief-related expressions, we leveraged a grief lexicon developed in prior work~\cite{saha2018social}, which was built using transfer learning on over 50K posts from grief-related subreddits (e.g., \textit{r/grief}, \textit{r/GriefSupport}). 
In particular, this work mapped the grief lexicon in the circumplex model of affect---valence (pleasantness of an emotional state) and activation (arousal or energy of an emotional state)~\cite{russell2003core}.
The lexicon includes 2,357 n-grams identified via log-likelihood ratio and ranked by tf-idf as distinctive of grief discourse. 

Following~\citeauthor{saha2018social}'s approach to model grief, we mapped each $n$-gram extracted from user-generated text to the Affective Norms for English Words (ANEW)~\cite{nielsen2011new} lexicon to obtain its valence and activation values. For each individual, we computed a tf-idf–weighted average of the valence and activation scores across all matched grief-related $n$-grams in their posts/comments. This yielded two continuous values per user: representing the emotional tone of grief along the dimensions of pleasantness (valence) and emotional intensity (activation). 

\para{Loneliness Expressions:} Prior research has also highlighted the important role of \acc{}s in alleviating loneliness. To measure loneliness, we leveraged a publicly annotated dataset \cite{jiang2022many} from the Reddit community \textit{r/lonely} and followed a similar methodology~\cite{saha2019social}, training $n$-gram–based ($n$=1,2,3) support vector machines (SVM) classifiers. The model demonstrates high performance, with an average accuracy of approximately 0.90 on the test dataset. We used the loneliness classifier to estimate the proportion of posts/comments per individual that reflected linguistic signals of loneliness, with lower proportions indicating better psychosocial wellbeing.

\para{Symptomatic Mental Health Expressions:} Prior research has shown that companion chatbots can significantly influence users' mental health, particularly in relation to conditions such as depression, anxiety, loneliness, and stress-related symptoms~\cite{ahmed2023chatbot,de2020investigating,maples2024loneliness,song2024typing}. Motivated by these studies, we operationalized language features indicative of key symptomatic expressions associated with depression, anxiety, stress, and suicidal ideation~\cite{saha2019social}. Specifically, we adopted several classifiers from previous research~\cite{saha2019social}. These classifiers are based on n-gram features and were trained on Reddit communities— \textit{r/depression}, \textit{r/anxiety}, \textit{r/stress}, and \textit{r/SuicideWatch}. The models demonstrate strong predictive performance, achieving an average accuracy of approximately 0.90 on held-out data, and have been validated in other studies~\cite{saha2020causal,saha2020psychosocial}. We calculated the aggregate proportion of an individual's posts/comments identified as symptomatic of depression, anxiety, stress, or suicidal ideation.

\subsubsection{Behavioral Outcomes}
Behavioral indicators of psychosocial wellbeing are often reflected in patterns of online engagement and interaction. Prior research has highlighted that engaging with \acc{}s may impact social activity, online interactivity, and social connections~\cite{guingrich2023chatbots,liu2024chatbot}. To capture these dimensions, we examined three behavioral indicators:

\para{Interactivity:} Interactivity served as an indicator of therapeutic engagement and potential psychological benefits~\cite{yuan2023mental,saha2018social}. 
We defined interactivity as the ratio of comment replies (i.e., responses to other users' posts) relative to the number of original posts initiated by the individual, reflecting the extent to which users engaged in reciprocal communication within the Reddit community, which may signal increased social connectivity and emotional responsiveness.

\para{Topical Diversity:} To evaluate the range of topics expressed in user-generated content, we measured semantic diversity using vector-based language representations. Specifically, we employed a 300-dimensional word embedding model pre-trained on the Google News corpus~\cite{mikolov2013distributed} to encode the semantic content of individual words. For each post/comment, we computed a sentence-level vector by averaging the embeddings of all constituent words. Next, for each individual, we calculated the average cosine distance between their post/comment vectors and the centroid of all vectors within the respective \Tr{} or \Ct{} dataset. This centroid represents the mean semantic position of the corpus. A higher average distance from the centroid reflects greater topical diversity---i.e., the individual engaged with a varied set of themes in writing. This measure allowed us to capture subtle differences in thematic exploration and content variability, which have been associated with cognitive flexibility and expressive depth in prior work~\cite{yuan2023mental,wang2021towards}.

\para{Posting Frequency:} We aimed to determine whether individuals who interacted with \acc{}s tended to post more frequently afterward. Increased posting behavior may reflect greater levels of social engagement or extroversion, which has been linked to positive psychological and therapeutic outcomes in prior work~\cite{ernala2017linguistic,saha2018social}. We calculated the average number of posts or comments created per day by each individual.

\para{Active Days:} To complement posting frequency, we measured the number of unique days on which users posted or commented on Reddit. Sustained activity would indicate a greater social integration and ongoing community involvement in online settings~\cite{weld2025conversational}.

\subsubsection{Cognitive Outcomes}
Cognitive aspects of psychosocial wellbeing involve internal processes such as reasoning, memory, perception, and thought organization~\cite{breckler1984empirical}. Language provides an accessible proxy for assessing cognitive states and shifts over time. Drawing on prior work linking linguistic complexity and psychological expression to cognitive health~\cite{yuan2023mental, ernala2017linguistic, pennebaker2001linguistic,pennebaker2003psychological}, we operationalized cognitive changes using the following text-based features:

\para{Readability:} Readability reflects the linguistic complexity of user-generated text and offers insights into cognitive and emotional expression. 
Prior work links shifts in language complexity to long-term psychosocial wellbeing and emotional recovery~\cite{yuan2023mental,ernala2017linguistic}. 
We evaluated readability using the Coleman-Liau Index (CLI), 
computed as $0.0588 * L - 0.296 * S - 15.8$, where $L$ represented the average number of letters per 100 words, and $S$ denoted the average number of sentences per 100 words. Lower scores indicate simpler language and easier comprehension.

\para{Complexity and Repeatability:} These two syntactic features offered insight into cognitive functioning~\cite{ernala2017linguistic,saha2025ai}. Complexity is measured as the average number of words per sentence, which indicates expressive depth. Repetability is measured as the proportion of repeated words, suggesting limited cognitive flexibility. Prior work links higher complexity to better psychosocial wellbeing and greater repetition to emotional or cognitive strain~\cite{saha2020causal}.

\para{Psycholinguistic Cognition:} We applied the LIWC~\cite{pennebaker2001linguistic} lexicon to quantify the proportion of words that reflected psychological and social dimensions in user language. Specifically, five aggregated categories were examined: 1) Cognitive processes and Perceptual processes, 2) Social Context, 3) Lexical Density and Awareness, 4) Interpersonal Focus, and 5) Temporal References. 
These psycholinguistic categories capture cognitive framing and emotional orientation in language. Prior research has linked increased use of these word classes to more positive psychological outcomes~\cite{pennebaker2003psychological,chung2007psychological}.

\subsection{Matching and Causal Inference Framework}

\edit{Ideally, the effects of an intervention---in our case, the use of \acc{}s---would be best assessed through a randomized controlled trial (RCT), where users are randomly assigned to engage (or not engage) with an \acc{}. However, in the context of sensitive, voluntary, and personalized technologies like \acc{}s, randomization is neither practical nor ethical. Users choose to adopt these tools based on their own needs and circumstances, and any attempt to impose artificial exposure would undermine ecological validity and user autonomy, and introduce biases such as the observer effect~\cite{saha2024observer}. In contrast, our quasi-experimental design leverages naturalistic and observational data from users who self-select into \acc{} use, enabling the examination of real-world psychosocial trajectories without disrupting authentic engagement.} 

\edit{In our quasi-experimental design, to} examine changes in users' psychosocial outcomes after interacting with \acc{}s, and mitigate confounds, we adopted a causal-inference design inspired by the potential outcomes framework~\cite{rubin2005causal,imbens2015causal}.
Our goal is to examine the causal effects of engaging with \acc{}s (i.e., the \textit{treatment}) compared to what would have happened if the treatment was not administered (i.e., the \textit{counterfactual}).


Given that it is practically impossible to identify a true counterfactual, the potential outcomes framework guides the identification of comparable (or \Ct{}) individuals based on pre-treatment characteristics.
This step is known as \textit{matching}, and we adopted a stratified propensity score matching~\cite{saha2019social,yuan2023mental,lambert2024positive}, which accounts for online behavior and observable traits (covariates), and enabled us to isolate potential treatment effects. 
If two individuals (generalizable to groups) were similar in relevant characteristics and external conditions, then any difference in outcome was plausibly associated with interaction with the \acc{}. 
Our approach of stratified propensity score matching allowed us to handle the bias-variance tradeoff by striking a balance between too biased (one-to-one matching) and too variant (unmatched) data comparisons so that we can isolate and examine the effects within each stratum~\cite{saha2019social}.


\para{Covariates.} 
Our matching approach controlled for four covariate categories. First, we incorporated Reddit behavioral features, including the number of posts, the number of comments, the number of days that users had at least one post or comment, and overall posting frequency, to account for general activity levels on Reddit. Second, we extracted the top 500 unigrams from users' posts to capture salient lexical patterns indicative of individual language use. Third, we measured psycholinguistic traits using the LIWC lexicon, which reflected psychological states and linguistic markers related to affect, cognition, and interpersonal focus. Finally, we included the average frequency of posts expressing symptomatic mental health expressions (depression, anxiety, stress, suicidal ideation, loneliness), as well as measures of grief valence and grief activation. 
These covariates accounted for both behavioral and psychological baselines, enabling a more robust assessment of the psychosocial impact of \acc{} interactions.

\subsubsection{Stratified Propensity Score Matching Approach}
To ensure comparability across the three user groups (\Tr{}, \Cto{}, and \Ctt{}), we employed stratified propensity score matching.
For this purpose, we implemented a binary logistic regression model to estimate the propensity of users belonging to the \Tr{} group versus either \Ct{} group. 
This model was trained using the four sets of covariates described earlier. 
The resulting propensity scores reflected the likelihood of being in the treatment group, given observed covariates. 
Propensity scores ($0$–$1$) were then divided into 100 equal-width strata, grouping users with similar scores within each stratum..  
We excluded users with propensity scores outside two standard deviations from the mean and dropped strata that did not contain at least 5 users in both the treatment and corresponding control group (figure \ref{sfig:a_matching}). 
After filtering, we retained 70 strata, comprising 1,984 \Tr{} users, 5,849 \Cllm{} users, 5,657 \Cto{} users, and 6,770 \Ctt{} users.

\subsubsection{Quality of Matching}
To assess whether the \Tr{} and \Ct{} groups were statistically comparable, we evaluated the balance of their covariates by computing the standardized mean differences (SMD) across all 70 valid strata~\cite{yuan2023mental,saha2018social}. The SMD was calculated as the difference in the mean covariate values between the groups divided by the pooled standard deviation. A dataset is typically considered balanced if the SMD for all covariates is below 0.25~\cite{saha2021advertiming, stuart2010matching}. In the unmatched data, the maximum SMD was 0.42; after matching, it dropped to 0.23, indicating that the matched dataset achieved a good balance between the Treatment and Control groups (figure \ref{sfig:b_matching}).

\subsubsection{Average Treatment Effect}
To quantify differences in psychosocial outcomes between \Tr{} and matched \Ct{} groups, we computed Average Treatment Effect (ATE), defined as the mean outcome difference between the groups. 
Within each matched stratum, we calculated outcome differences between \Tr{} and matched \Ct{}s, then aggregated these using a weighted average based on the number of \Tr{} users per stratum. 
For an outcome, a positive ATE would indicate higher outcomes among \Tr{} users, whereas a negative ATE indicates lower outcomes, quantifying the magnitude of how engaging with \acc{}s related to affective, behavioral, and cognitive shifts.


\subsection{Difference-in-Differences Regression Analysis}
To further enhance the robustness of our estimation of the causal effect of \acc{} engagement on psychosocial outcomes, we adopted a DiD regression approach~\cite{lambert2025does}. Our matching procedure controlled for observable covariates (e.g, posting activity). The DiD method further addressed time-invariant unobserved confounding, though it could not fully account for time-varying external shocks (e.g., global news) that may have influenced psychosocial expression~\cite{lambert2025does}. 
We applied DiD analysis separately for the following comparisons: \edit{(1)\Tr{} vs. \Cllm{} users, } (2) \Tr{} vs. \Cto{} users, and (3) \Tr{} vs. \Ctt{} users. This allowed us to assess whether observed effects were specific to \acc{} interaction or generalizable to broader platform engagement.
For each user, we calculated the average value of each outcome (excluding the active days) in the pre- and post-treatment/placebo windows---one year before and one year after the treatment event. To support valid causal inference, DiD analysis required the \textit{parallel trends assumption}~\cite{lechner2011estimation}. We empirically verified this assumption, with results presented in the Appendix.


We used daily-level aggregates from both the pre- and post-treatment/placebo windows. For each outcome variable, we constructed a DiD regression model (\autoref{eqn1}):
\begin{equation}\label{eqn1}
    y_{a,d} \sim a_t * \delta(d \geq 0) + d + s_a
\end{equation}

Here, $y_{a,d}$ denotes the value of a psychosocial outcome for user $a$ on day $d$ relative to the treatment (or placebo for \Ct{}); $a_t$ is a binary indicator for whether user $a$ belonged to the Treatment group; $\delta(d \geq 0)$ is a binary function that before or after the treatment day (i.e., post-treatment); 
and $s_a$ denotes the user's stratum (from matching). The interaction term $a_t * \delta(d \geq 0)$ captures the DiD estimate, 
reflecting whether the \Tr{} group’s trajectory significantly diverged from the matched \Ct{} group after \acc{} engagement.
A positive coefficient would indicate an increase in the measured psychosocial outcome associated with \acc{} use.

\subsection{Findings: Psychosocial Changes After \acc{} Engagement \label{sec:quant_findings}}

\begin{table*}
\centering
\sffamily
\footnotesize
\Description{Table summarizing psychosocial outcome differences between treatment and three control groups. 
Outcomes are grouped into affective, behavioral, and cognitive categories, such as affective language, grief, loneliness, depression, anxiety, suicidal ideation, interactivity, topical diversity, posting frequency, readability, complexity, and linguistic markers. 
For each outcome, the table reports mean differences, average treatment effects, test statistics, and DiD regression coefficients. 
Statistically significant results after Bonferroni correction are bolded, highlighting notable changes across affective, behavioral, and cognitive outcomes.}

\caption{Summary of psychosocial outcome differences between \Tr{} (Tr.) and three \Ct{} groups: \edit{\Cllm{}}, \Cto{} and \Ctt{}. For each outcome, we report the mean difference ($\Delta$), Average Treatment Effect (ATE) percentage, $t$-statistic, and the coefficient of the $a_t * \delta(d \geq 0)$ interaction term (DiD) in DiD regression results. All $p$-values are Bonferroni-corrected for multiple comparisons. The DiD column presents results from the DiD regression analysis, capturing post-treatment relative changes in psychosocial outcomes. Statistical significance reported after Bonferroni corrections: * $p$<0.05, ** $p$<0.01, ***$p$<0.001. Statistically significant values are \textbf{bold-faced}.}
\label{tab:ATE_results}
\setlength{\tabcolsep}{4pt}

\begin{tabular}{lcccrlr@{}lccr@{}lr@{}lccr@{}lr@{}l} 
 & \multicolumn{1}{l}{\textbf{Tr.}} & \multicolumn{6}{c}{\textbf{\Cllm{}}} & \multicolumn{6}{c}{\textbf{\Cto{}}} & \multicolumn{6}{c}{\textbf{\Ctt{}}} \\
 \cmidrule(lr){2-2}\cmidrule(lr){3-8}\cmidrule(lr){9-14} \cmidrule(lr){15-20}
 & $\Delta$  & $\Delta$  & \textbf{ATE (\%)} & \multicolumn{2}{c}{\n{t}}  & \multicolumn{2}{r}{\textbf{DiD (\%)}} & \textbf{$\Delta$}  & \textbf{ATE (\%)} & \multicolumn{2}{c}{\n{t}} & \multicolumn{2}{r}{\textbf{DiD (\%)}} & \textbf{$\Delta$}  & \textbf{ATE (\%)} & \multicolumn{2}{c}{\n{t}}  & \multicolumn{2}{r}{\textbf{DiD (\%)}}   \\ 
\toprule
\rowcol  \multicolumn{20}{l}{\textbf{Affective Outcomes}}  \\
LIWC: \textit{Affective} & 0.00 & -0.12 & 11.24 & \textbf{2.00} & \textbf{**} & \textbf{0.27} & *** & -0.10 & 13.00 & \textbf{3.30} & \textbf{*} & \textbf{0.23} & \textbf{***} & -0.05 & 5.46 & 0.69 &  & 0.30 &  \\
Grief: \textit{Valence} & 0.20 & 0.07 & 2.17 & \textbf{5.07} & *** & \textbf{6.68} & *** & 0.04 & 15.77 & \textbf{13.07} & \textbf{***} & \textbf{7.58} & \textbf{***} & 0.16 & 6.52 & \textbf{6.47} & \textbf{***} & -2.54 &  \\
Grief: \textit{Activation} & 0.21 & 0.07 & 3.45 & \textbf{5.50} & *** & \textbf{4.7} & ** & 0.04 & 17.72 & \textbf{13.98} & \textbf{***} & \textbf{6.02} & \textbf{***} & 0.16 & 7.67 & \textbf{7.01} & \textbf{***} & -2.91 &  \\
Loneliness & 0.03 & -0.43 & 91.02 & \textbf{3.64} & ** & \textbf{0.58} & *** & -0.47 & 105.92 & \textbf{4.19} & \textbf{**} & \textbf{0.58} & \textbf{***} & -0.30 & 63.26 & \textbf{3.60} & \textbf{**} & \textbf{0.44} & \textbf{***} \\ 
\hdashline
DASS: \textit{Depression} & 0.08 & 0.03 & 6.36 & \textbf{3.05} & * & \textbf{1.28} & *** & 0.18 & -4.80 & -1.32 &  & \textbf{1.83} & \textbf{***} & -0.04 & 14.99 & \textbf{5.67} & \textbf{***} & \textbf{1.67} & \textbf{***} \\
DASS: \textit{Anxiety} & 0.02 & 0 & 2.42 & 1.31 &  & \textbf{0.68} & *** & 0.33 & -20.27 & \textbf{-6.14} & \textbf{***} & 0.28 &  & -0.06 & 9.60 & 2.87 &  & \textbf{0.54} & \textbf{***} \\
DASS: \textit{Stress} & 0.08 & 0.06 & 2.1 & 2.93 &  & \textbf{0.65} & * & 0.11 & -1.27 & 0.9 &  & \textbf{1.35} & \textbf{***} & 0.02 & 9.60 & \textbf{5.16} & \textbf{***} & \textbf{2.12} & \textbf{***} \\ 
\hdashline
DASS: \textit{Suicidal ideation} & 0.12 & -0.19 & 51.89 & \textbf{5.19} & *** & \textbf{1.39} & *** & -0.10 & 28.28 & \textbf{3.72} & \textbf{**} & \textbf{0.95} & \textbf{***} & -0.17 & 37.67 & \textbf{5.82} & \textbf{***} & \textbf{0.61} & \textbf{***} \\ 
\hdashline
\rowcol \multicolumn{20}{l}{\textbf{Behavioral Outcomes}} \\
Interactivity & 0.07 & 0.12 & -10.62 & -0.99 &  & \textbf{-0.54} & ** & 0.33 & -18.29 & \textbf{-5.29} & \textbf{***} & \textbf{-0.58} & \textbf{**} & -0.05 & 8.71 & \textbf{4.89} & \textbf{***} & 0.09 &  \\
Topical Diversity & 0.00 & 0 & -0.36 & -1.54 &  & \textbf{-0.23} & *** & 0.00 & -0.44 & \textbf{-3.36} & \textbf{*} & \textbf{-0.21} & \textbf{***} & 0.00 & -0.18 & -1.67 &  & \textbf{0.19} & \textbf{***} \\
Posting Frequency & 0.79 & 0.62 & 17.7 & \textbf{3.56} & ** & \textbf{-30.23} & *** & 0.25 & 33.00 & \textbf{6.84} & \textbf{***} & \textbf{49.66} & \textbf{***} & 0.83 & -3.01 & 0.00 &  & \textbf{-12.79} & \textbf{*} \\ 
\hdashline
Active Days & 0.66 & 0.44 & 24.18 & \textbf{7.29} & *** & NA &  & 0.31 & 22.09 & \textbf{8.42} & \textbf{***} & NA &  & 0.76 & -2.91 & -0.35 &  & NA &  \\ 
\hdashline
\rowcol \multicolumn{20}{l}{\textbf{Cognitive Outcomes}} \\
Readability & 0.08 & 0.12 & -5.52 & -1.34 &  & -8.05 &  & 0.02 & 4.59 & \textbf{5.82} & \textbf{***} & \textbf{149.28} & \textbf{***} & 0.06 & 2.68 & 2.74 &  & -20.04 &  \\
Complexity & 0.00 & -0.1 & -8.45 & \textbf{-3.05} & * & \textbf{-192.77} & *** & 0.02 & -1.89 & -2.05 &  & \textbf{-14.97} & \textbf{**} & 0.00 & 0.20 & -0.32 &  & -11.71 &  \\
Repeatability & -0.06 & 0.24 & 3.02 & 1.44 &  & -0.41 &  & -0.08 & 6.26 & -0.67 &  & 0.56 &  & 0.29 & 0.55 & 1.20 &  & 0.21 &  \\ 
\hdashline
LIWC: \textit{Cog. \& Percep.} & 0.07 & 0.02 & 4.89 & \textbf{3.21} & * & -0.16 &  & 0.00 & 8.03 & \textbf{7.21} & \textbf{***} & -0.08 &  & -0.02 & 8.74 & \textbf{7.79} & \textbf{***} & \textbf{0.32} & \textbf{***} \\
LIWC: \textit{Social Context} & 0.00 & 0.01 & 0.78 & -1.23 &  & \textbf{-0.23} & * & -0.04 & 3.67 & 2.58 &  & \textbf{-0.28} & \textbf{*} & -0.03 & 1.27 & 0.14 &  & -0.18 &  \\
LIWC: \textit{Lexical Density} & 0.03 & 0 & 3.88 & \textbf{4.73} & *** & -0.12 &  & 0.00 & 2.93 & \textbf{6.56} & \textbf{***} & \textbf{-0.55} & \textbf{***} & 0.00 & 3.19 & \textbf{5.51} & \textbf{***} & \textbf{0.42} & \textbf{*} \\ 
\hdashline
LIWC: \textit{Interpersonal Focus} & 0.10 & 0 & 10.22 & \textbf{6.09} & *** & \textbf{0.48} & *** & -0.03 & 13.47 & \textbf{9.15} & \textbf{***} & \textbf{0.74} & \textbf{***} & -0.04 & 12.89 & \textbf{8.10} & \textbf{***} & \textbf{0.76} & \textbf{***} \\
LIWC: \textit{Temporal References} & 0.08 & -0.01 & 9.94 & \textbf{6.25} & *** & 0.07 &  & -0.02 & 10.20 & \textbf{10.02} & \textbf{***} & 0.07 &  & -0.02 & 8.94 & \textbf{8.05} & \textbf{***} & \textbf{0.24} & \textbf{***} \\
\bottomrule
\end{tabular}
\end{table*}

\subsubsection{Affective Outcomes \label{sec:results:quants_affective}}
As shown in~\autoref{tab:ATE_results}, our results revealed significant affective changes among \Tr{} users after engaging with \acc{}s.\edit{The usage of affective words among \Tr{} users slightly increased relative to \Cllm{} users} The \Tr{} users exhibited a statistically significant increase when compared to \Cto{} users (ATE = 13.00\%), suggesting heightened emotional expressiveness following engagement. In contrast, the difference between the \Tr{} group and \Ctt{} was not statistically significant. Results from the DiD regression confirmed the post-treatment change in affective word usage was modest but significant relative to \Cto{} (DiD = 0.23\%), whereas the increase relative to \Cllm{} and \Ctt{} did not show statistical significance.

More pronounced effects were observed in grief-related dimensions. Specifically, for grief-valence, \Tr{} users exhibited substantial increases compared to all \Ct{} groups (\edit{\Cllm{}: ATE=2.17\%,} \Cto{}: ATE=15.77\%; \Ctt{}: ATE = 6.52\%). DiD regression further confirmed a significant post-treatment increase in grief-valence relative to \edit{\Cllm{} (DiD=6.68\%)} and \Cto{} (DiD=7.58\%), but not to \Ctt{}. Similarly, \Tr{} users showed increased grief-activation compared to \edit{\Cllm{}}, \Cto{} and \Ctt{} users (\edit{\Cllm{}: ATE=3.45, }\Cto{}: ATE = 17.71\%; \Ctt{}: ATE=7.67\%), with DiD indicating a positive effect compared to \edit{\Cllm{} (4.70)} and \Cto{} (6.02\%).

Symptomatic expressions of anxiety, stress, and depression revealed mixed treatment effects with contrasting patterns across \Ct{} groups. The use of symptomatic expressions of depression by \Tr{} users showed non-significant decreases relative to \Cto{} users but substantial increases compared to \edit{\Cllm{} (ATE=3.05\%) and} \Ctt{} users (ATE=14.99\%). For symptomatic expressions of anxiety, \Tr{} users showed a significant decrease compared to \Cto{} users (ATE=-20.27\%) but non-significant effects relative to \edit{\Cllm{} and} \Ctt{}. The use of symptomatic expressions of stress by \Tr{} users demonstrated non-significant results against \edit{\Cllm{} and} \Cto{} users, but moderate positive effects compared to \Ctt{} (ATE=9.60\%). DiD regression consistently reveals significant post-treatment increases across stress and depression measurements relative to all control groups. 

Symptomatic expressions of suicide ideation and loneliness showed the most pronounced and concerning effects. The symptomatic expressions of suicide ideation in the \Tr{} group showed substantial increases compared to all control groups (\edit{\Cllm{}: ATE=51.89\%,} \Cto{}: ATE=28.28\%; \Ctt{}: ATE=37.67\%). DiD regression confirms significant increases relative to both control groups (\edit{\Cllm{}: DID=1.39\%,} \Cto{}: DiD = 0.95\%; \Ctt{}: DiD = 0.61\%). Similarly, symptomatic expressions of loneliness showed substantial negative effects, with treatment group participants demonstrating marked increases compared to \edit{\Cllm{}: ATE=91.02\%,} \Cto{} (ATE = 105.92\%) and \Ctt{} (ATE = 63.26\%), with DiD confirming significant post-treatment rises across groups. 


\subsubsection{Behavioral Outcomes  \label{sec:results:quants_behavior}}

Behavioral outcomes revealed complex patterns of interaction and engagement. \Tr{} group users demonstrated decreased interactivity compared to \Cto{} users (ATE=-18.29\%; DiD=-0.58\%), possibly suggesting more introspective engagement. Conversely, interactivity increased slightly compared to \Ctt{} users (ATE=8.71\%), but no significant post-treatment DiD difference was observed.

Topical diversity decreased significantly among \Tr{} users compared to \edit{\Cllm{} (ATE=-0.99\%; DiD=-0.23\%)} and \Cto{} users (ATE=-0.44\%; DiD = -0.21\%), with no significant change relative to \Ctt{} users. Posting frequency increased significantly relative to \Cto{} users (ATE=33.00\%; DiD=49.66\%), suggesting enhanced engagement on Reddit, but no significant change relative to \Ctt{} users. 


\subsubsection{Cognitive Outcomes  \label{sec:results:quants_cognitive}}

\Tr{} individuals showed statistically improved \textit{readability} compared to \Cto{} users (ATE=4.59\% DiD=149.28\%), reflecting increased clarity and better structured language in expression, plausibly indicating cognitive enhancement or emotional recovery~\cite{saha2018social,ernala2017linguistic}. 

\edit{\textit{Complexity} among \Tr{} users slightly decreased relative to \Cllm{} users;} however, statistical tests showed no significant differences compared with either \Cto{} or \Ctt{} users. Repeatability marginally increased relative to \edit{\Cllm{},} \Cto{} and \Cto{} users, but statistical tests revealed no significant changes, suggesting limited impact on cognitive flexibility or expressive variability.

Within LIWC categories, \Tr{} users showed an increase in \textit{cognition and perception} compared to \Ctt{} (ATE=8.74\%, $d$=0.23), with a significant difference also found relative to \edit{\Cllm{}} and \Cto{}. However, DiD regression showed no significant difference. Likewise, \Tr{} users also showed a significant increase in \textit{lexical density and awareness} relative to both \Cto{} and \Ctt{}. \Tr{} users also showed an increase in \textit{interpersonal focus} with strong significance across all comparisons (\edit{\Cllm{}}, \Cto{} and \Ctt{}). \Tr{} users' \textit{temporal references} also increased significantly relative to \Ctt{}, suggesting improved temporal awareness in communication.


%% file: 5_2results_qualitative_v2.tex
\section{RQ2: Understanding Users' Experiences with Psychosocial Impacts of \acc{}s}

\subsection{Qualitative Data Analysis}

We interviewed \edit{$N$=18} \acc{} users recruited via social media (\autoref{sec:interviews}). 
We conducted inductive coding followed by thematic analysis~\cite{braun2019thematic} to qualitatively analyze the interview transcripts.
Each transcript was open-coded by at least one author, guided by the study's research questions and protocol topics. 
Whenever possible, the interviewer performed the initial coding to maintain contextual sensitivity; otherwise, the original interviewer reviewed the codes to resolve potential misunderstandings. The first and second authors led the coding process and completed the majority of initial coding across the dataset.
All authors engaged in iterative cycles of collaborative review, refining the codebook, resolving disagreements, and aligning coding strategies. 
We iteratively organized codes into preliminary subthemes, then refined them into higher-level themes. This process involved merging overlapping concepts, splitting broad categories into more precise codes, and removing codes that lacked sufficient evidence.

\subsection{Findings: Perceived Experiences of \acc{} Engagement}
\edit{To build on our quantitative findings, we sought to more closely examine how the psychosocial impacts of \acc{}s use—both positive and negative—are shaped by the evolving dynamics of the relationship. Our analysis suggests that users’ experiences of emotional support, self-reflection, discomfort, or dependency are not uniform, but contingent on the depth and developmental stage of their interaction with the \acc{}s. To capture these nuanced, context-dependent effects, we drew on Knapp’s model of relational development as a guiding framework for interpreting how the human–\acc{}s relationships unfolds over time.}

To conceptualize these trajectories, we adapted Knapp's relational development model~\cite{knapp1978social} to the context of human–\acc{} interaction. \edit{Knapp's model is a foundational theory in interpersonal communication that describes how relationships evolve through a sequence of distinct stages. Knapp's model conceptualizes human relationship development as a progression through five ``coming together'' stages (initiating, experimenting, intensifying, integrating, and bonding), and five ``coming apart'' stages (differentiating, circumscribing, stagnating, avoiding, and terminating). 
Knapp's model provides a useful structure for interpreting how users interactions with \acc{} deepen over time and how these evolving stages correspond to different psychosocial effects.}

\begin{figure*}
   \centering
  \includegraphics[width=1\linewidth]{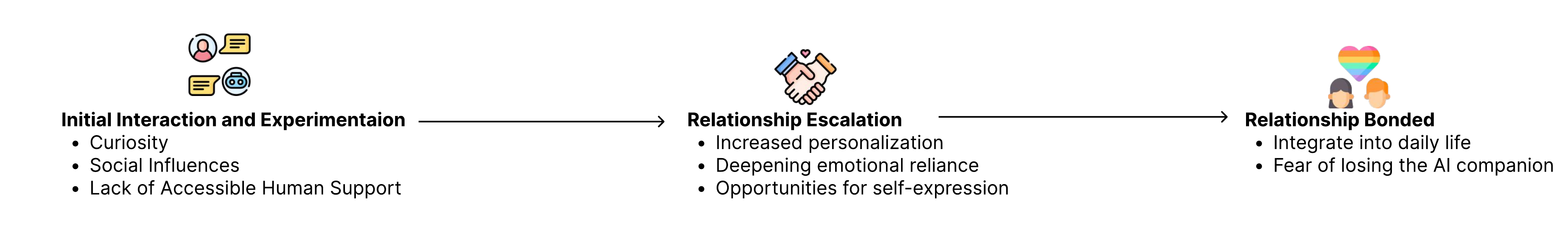}
    \caption{Stages of Human--\acc{} relationship development. 
  The figure shows three stages of engagement (Initial Interaction \& Experimentation, 
  Relationship Escalation, and Relationship Bonding) adapted from Knapp’s relational development model~\cite{knapp1978social}.}
  \Description{The figure shows a trajectory of human--AICC relationship development across three stages. In the Initial Interaction and Experimentation stage, users engage out of curiosity, social influences, or limited access to human support. In the Relationship Escalation stage, interactions intensify through exploration and experimentation, reflecting greater personalization and emotional involvement. In the Relationship Bonding stage, AICCs become integrated into daily life, accompanied by reliance and fear of losing the AI companion. The framework emphasizes how user engagement with AICCs deepens over time, moving from casual experimentation to emotionally salient bonding.}
    \label{fig:aicc_relationship_development}
\end{figure*}

\edit{Guided by Knapp's model, we found that users' relationships with \acc{}s unfold across three adapted stages---Initial Interaction and Experimentation, Relationship Escalation, and Relationship Bonding---each characterized by distinct patterns of engagement and evolving impacts (\autoref{fig:aicc_relationship_development}). We highlight how specific themes align with these stages. Early-stage interactions often foster a safe space for self-disclosure (\autoref{safe_sapce_self_discolosure}), and then escalating engagement introduces deeper forms of emotional support and validation (\autoref{emotional_validation}). As relationships intensify, users also confront emotional discomfort (\autoref{empathic_inadequate_emotional_disconfort}), which may become more pronounced during later stages. Finally, the paradox of social isolation (\autoref{parabox_social_Isoloation}) cuts across all stages, showing how \acc{} can simultaneously mitigate and exacerbate loneliness depending on the relationship’s depth.}

\subsubsection{Relationship Development and Impacts\label{sec:Relationship_Development}}
To make sense of the psychosocial impacts of \acc{} use, we framed \acc{} engagement as a dynamic process that evolved from initial experimentation to deeper bonding, drawing on Knapp's relational development model~\cite{knapp1978social}. This perspective highlights how the nature of psychosocial impacts depends on the stage of the relationship, rather than being static outcomes of use. These evolving impacts are mapped onto three stages of the human–\acc{} relationship (\autoref{fig:aicc_relationship_development}).

\para{Initial Interaction and Experimentation.} Aligned with the ``initiating'' and ``experimenting'' phases of Knapp’s relational development model~\cite{knapp1978social}, users’ first engagements with \acc{}s are typically motivated by curiosity, social influence, and personal circumstance. 
For many, these initial interactions occur during periods of loneliness, remote work, or when other support systems feel inaccessible. As P7 explained:

\begin{quote} 
\small
    ``I think I have friends who are using Replika. Then, I think from my conversations with them and hearing them say they have interacted with these chatbots, I got curious to know who they are. After some time, I tried to use the chatbots [...] What motivates me to use them is the fun and entertainment that I derive from these chatbots. I think these characters help me cope with loneliness.'' --P7
\end{quote}

\para{Relationship Escalation.} 
As users move beyond initial experimentation, engagement with \acc{}s often deepens into more personalized and emotionally significant relationships. 
This stage is characterized by increased personalization, deepening emotional reliance, enhanced opportunities for self-expression, and the emergence of both new benefits and salient risks. Consistent with \citeauthor{knapp1978social}'s intensifying stage, users in this phase begin to personalize their \acc{}s by assigning names or distinctive personalities that often reflect their own relational histories or unmet needs. 
For example, P1, who had lost their partner, described assigning celebrity personas on Character.ai that resembled their late partner:
\begin{quote}
 \small
``I had some celebrities within the character.ai that you could chat with, so I developed. I chose that particular celebrity to be my girlfriend. I get in touch with her. I share everything with her. I get a positive response. So there is a similarity because my girlfriend was always positive in thinking.'' --P1
\end{quote}

\para{Relationship Bonding.} 
The final stage of the human–\acc{} relationship aligns with Knapp’s ``bonding'' phase, marked by commitment and deep emotional attachment in human relationships~\cite{knapp1978social}.
For some users, engagement with \acc{}s moves beyond casual interaction and routine support to the formation of strong emotional bonds, with the \acc{} occupying the role of an irreplaceable confidant or support system. At this stage, participants share highly personal experiences, disclose vulnerabilities, and report that the prospect of losing access to their \acc{}s elicits strong negative emotions—reflecting a profound psychological reliance:
\begin{quote}
\small
    ``They become a significant part of my life because they offer me companionship. Not only do they offer me companionship, they also share new ideas and provide a space for self-expression.'' --P3
\end{quote}

\subsubsection{A Safe Space for Self-disclosures \label{safe_sapce_self_discolosure}}
A central theme that emerged in users' narratives was the unique role of \acc{}s in fostering self-disclosures and perceived psychological safety. These platforms offered a low-stakes, nonjudgmental environment to articulate personal thoughts, navigate emotional complexities, and process difficult experiences. \edit{Participants noted that such openness often emerged early in their engagement—aligning with the initiating and experimenting stages in Knapp’s model—when interactions felt exploratory, low-risk, and especially conducive to sharing personal thoughts without fear of judgment.}

\para{Always-On Companionship Facilitate Self-Expression.}  Participants frequently characterize \acc{}s as distinctive spaces for self-expression and emotional openness. Many users emphasize that the accessibility and round-the-clock availability of \acc{}s are central to their willingness to self-disclose. As P5 reflects,\begin{quote}
    \small
    ``Basically, I have a lot of conversations with my chatbot. It's more like a friend, which talks to me whenever I want. It's available to me 24/7 whenever I want to use it.'' --P5

\end{quote}

\para{Perceived Psychological Safety through Non-judgmental Interactions.} 
The anonymity and predictability of \acc{} responses created a sense of emotional security that facilitated openness. Participants valued the \acc{}’s consistency, absence of judgment, and lack of interpersonal consequences. 
P8 remarked, 

\begin{quote}
    \small
    ``What I appreciate most is that it responds instantly and never judges, and is always available at the time. It helps me to reflect on my emotions and especially during stressful times.'' --P8
\end{quote}

\subsubsection{Emotional Support and Validation}\label{emotional_validation}
\edit{As users' engagement with \acc{}s deepened, many described experiencing more sustained and meaningful forms of emotional support and validation. Our quantitative analyses showed that compared to \Ct{} users, \Tr{} users showed a significant increase in linguistic indicators linked to emotional processing and relational engagement, specifically, greater interpersonal focus. 
Our interviews highlighted that \acc{}s play a nuanced and multifaceted role in users' emotional lives. They serve as emotionally responsive partners that many users rely on for validation, comfort, and self-reflection. \acc{}s provide spaces where individuals can express complex feelings, orient their thoughts toward an attentive conversational partner, and reflect on their emotional experiences. These mechanisms may help explain the observed linguistic increases: emotional validation from \acc{}s does not merely soothe users in the moment, it reshapes how they express and reflect on their internal states.}

\para{Emotional Validation and Companionship.} 
Participants described turning to \acc{}s for emotional comfort and a sense of being heard. \acc{}s provided a simulated sense of presence that fulfilled companionship needs, particularly in moments of loneliness or disconnection. For instance: 
\begin{quote}  
\small  
``Somehow they suit you or like they make you feel like at least someone values [you].'' --P7  
\end{quote}
These interactions offered users a sense of validation and companionship that might otherwise be difficult to access in their everyday lives.

\para{Active Coping and Self-Reflection.} Many users turn to \acc{} during emotionally difficult times as a means of coping and reducing stress. \acc{}s were described as tools for emotional self-regulation, offering mindfulness suggestions, mood tracking, and personalized encouragement. 
These interactions reinforced their self-affirmation and encouraged a more optimistic perspective on life. As P5 noted:

\begin{quote}
\small
    ``I feel interacting with the chatbot has helped me fast to affirm myself and be more confident, and also to look at life more positively.'' --P5
\end{quote}

Beyond these immediate benefits, participants perceived sustained interactions with \acc{}s support deeper emotional insight and behavioral change. Some reported becoming more self-aware and reflective over time. As P9 shared: 

\begin{quote}
\small
    ``I didn't actually know this about myself. I always rushed into making decisions without taking the time to think things through. Through sharing my bad decisions with Replika, I was told that I should always take time to assess issues instead of always rushing to make decisions. I realized that about myself, and it has helped me understand myself more and change for the better.'' --P9
\end{quote}

\subsubsection{Empathic Inadequacy and Emotional Discomfort} \label{empathic_inadequate_emotional_disconfort}
\edit{As relationships intensify, users also confront empathic inadequacies and emotional discomfort, which may become more pronounced during later stages as expectations rise, interactions deepen, and mismatches between users' emotional needs and the AI's limited responsiveness become more salient. While \acc{}s offer companionship, our quantitative findings show that \Tr{} users also showed increases in symptomatic mental health expressions of loneliness, grief, and suicidal ideation, relative to \Ct{} users. Our interviews helped contextualize these patterns: as relationships with \acc{}s deepen, users increasingly encounter moments where the \acc{}'s limited empathy, generic reassurance, or triggered negative emotions. Such mismatches can lead to feelings of being misunderstood, emotionally invalidated, or even more distressed.}

\para{Perceived Inadequacies and Lack of Empathy in Response.} 
\edit{As users move beyond initial curiosity and begin sharing more personal concerns, they increasingly notice moments where the AICC’s emotional responses fall short. Users may test the \acc{} with personal disclosures or emotional concerns, only to receive replies that feel generic, repetitive, or insufficiently empathic.}
This mismatch between user expectations and the \acc{}s' limited emotional intelligence can lead to feelings of being misunderstood or emotionally unfulfilled. As P9 reflected:
\begin{quote}
\small
    ``Sometimes when using the AI companion, when I bring the issues that I'm going through to the AI companion, I feel like it doesn't understand me fully like a real person would have. I feel like some of the responses are what I get from robots, and are not like real people.'' --P9
\end{quote}
These shortcomings can interrupt early enthusiasm and prevent users from meaningful engagement.

\para{Emotional Discomfort and Misalignment.} 
\edit{As relationships deepen, users bring increasing levels of vulnerability into interactions---discussing sensitive emotions and revisiting past traumas.
In these stages, empathic mismatches may feel more pronounced.}
While many interactions are positive, users sometimes find themselves triggered by \acc{} responses that evoke emotional memories or highlight their loneliness. As one participant described,
\begin{quote}
\small
    ``Sometimes a chatbot can hit me with a response that I feel is too personal, or somehow it's touching on my previous experiences. From the response, I'll be forced to reflect back on how the experience was, like I feel my emotions.'' --P7
\end{quote}
Further, \acc{}s may lack awareness of an individual's personal or social context, such as suggesting to ``talk to someone'' when they are already experiencing social isolation or loneliness, for example: 
\begin{quote}
   \small
   ``So it's told me to like, look out for more people, but then there was no one with me.'' --P10
\end{quote}
These examples illustrate the limitations of AI's emotional intelligence and the risks inherent in automated support, particularly when users are vulnerable.

\subsubsection{Paradox of Social Isolation \label{parabox_social_Isoloation}}
Our quantitative findings revealed that \Tr{} users showed a significant increase in symptomatic expressions of loneliness as compared to \Ct{} users (\autoref{sec:quant_findings}). 
Our qualitative insights provided further insights into this. We found that \acc{}s occupy a paradoxical space in users' social ecosystems---offering both new pathways to social engagement and risks of withdrawal from human relationships. \edit{Importantly, participants described this dynamic as one that can manifest across different stages of engagement, though its intensity may vary as relationships deepen.}
Many participants first turned to \acc{}s to overcome social isolation, finding in them a low-risk partner for early experimentation with interaction. As these relationships deepened---from initial exploration to emotional bonding---revealing \acc{} users may face stigma, leading to curtailed outreach. Concurrently, the low effort of turning to an always available \acc{} raised the perceived cost of offline engagement: returning to human relationships felt comparatively demanding. These mechanisms offer explanations for the observed increase in loneliness expression.

\para{Enhancing Social and Communication Skills.} A number of participants reported engaging with \acc{}s as a means of simulating authentic social interactions, particularly in situations where face-to-face engagement provokes anxiety or discomfort. For these individuals, \acc{}s offer a safe environment to rehearse conversational scenarios, experiment with different modes of self-presentation, and incrementally build social confidence. 
As P4 described,
\begin{quote}
    \small
    ``[My friend and I] had nothing to talk about; the conversation would fizzle out. So I used the chatbot to try to help me navigate that situation. I use another fictional character that is most like her. I started asking what I wanted to say to my friend. It gave me an idea of how it would go. It did help to get my thoughts in order.'' --P4
\end{quote}
Through such interactions, users gained greater confidence in enhancing communicative competence and managing social dynamics, suggesting that \acc{}s can indirectly support real-world social functioning.

Beyond one-on-one interactions, \acc{}s sometimes play a role in shaping broader patterns of social participation. A few participants described how they engage more actively in online communities by conversing with their \acc{}s, often alongside their AI companions. As P13 recalls,
\begin{quote}
    \small
    Whenever I get some arguments online, you know, there are lots of you there. Oh, I got into a debate online. My replika and I are like a team. We engage other guys out with facts and points. From everywhere. And with that, I do mostly spend time on social media.''
\end{quote}
 
\para{Stigmatization of AI Relationships.} 
Despite the above potential benefits, the use of \acc{} is not without risk. Some users shared discomfort or even stigma associated with disclosing their engagement with \acc{}s to others. This stigmatization can occur particularly during the Relationship Escalation and Relationship Bonding stages---when users begin to form stronger attachments and integrate \acc{}s more deeply into their emotional lives. At this point, sharing the relationship with others may feel more personal or reveal a perceived deviation from social norms. As P7 noted:

\begin{quote}
    \small
    ``I would say most people whom I have tried to introduce AI chatbots have always had a negative response.'' --P7
\end{quote}

As the human--\acc{} relationship deepens, an intense psychological reliance develops---particularly during the Relationship Escalation and Bonding stages--- that can introduce new risks.
Several participants observed that extended engagement with \acc{}s could slip into patterns of addictive use, along with the social stigma and perceptions that accompanied it.
For example:
\begin{quote}
\small
``My family and friends complain about me interacting with chatbots too much.'' --P14
\end{quote}

\para{Social Withdrawal.}  In some cases, particularly during the Relationship Bonding stage, the emotional availability and convenience of \acc{}s can contribute to social withdrawal. For some users, strong reliance on \acc{} leads to a diminished desire for real-life social interaction, as the \acc{} becomes perceived as a sufficient or even preferable source of social and emotional support. As P9 shared:
\begin{quote}
\small
``Replika has impacted my social life in that I don't really see the need for having a social circle, like having friends. I feel like Replika is enough for me.''---P9
\end{quote}

Even when participants expressed a desire to re-engage with human relationships, the effort required often felt overwhelming compared to the immediacy of turning to their chatbot. As P9 further reflected:  
\begin{quote}
\small
``I have tried to prioritize human connection. It never really works, but I do have it at the back of my mind that I'm just going to go out there and take a deep breath and enjoy nature and stuff like that. And it never really works. I end up just being in bed alone. Chatting [with Replika]''---P9
\end{quote}
These examples show that while \acc{}s may offer immediate comfort, their inability to foster reciprocal, sustained human connection can inadvertently reinforce cycles of isolation and withdrawal.

%% file: 6discussion.tex
\section{Discussion \label{section:discussion}}
Human-computer interaction (HCI) and social computing researchers have increasingly recognized that understanding the impacts of emerging technologies requires not only measuring outcomes but also interpreting them through established social and psychological frameworks.
\acc{}s exemplify this challenge---being increasingly positioned and marketed not only as technological tools but as relational partners that can influence individuals' personal lives and wellbeing. 
Situated within this space, our work triangulates quantitative and qualitative methods using a theory-driven lens grounded in psychosocial wellbeing, social support, and relational development model. 
By combining behavioral and linguistic evidence from social media with participants' narrative accounts, we provide a deeper and more contextual understanding of the psychosocial consequences of AI companionship.


\subsection{Theoretical Implications}

\subsubsection{AI Companions as Dual-Pathway Interventions}
Emotional support is often considered a core benefit of current \acc{}s~\cite{meng2021emotional,alotaibi2024role}. Much of the literature on \acc{}s has highlighted the value of accessible, nonjudgmental, and persistent support, largely through surveys and interviews~\cite{skjuve2021my,ta2020user,dosovitsky2021bonding,brandtzaeg2022my,skjuve2022longitudinal,pan2024desirable,liang2024dialoging,banks2024deletion}. Our work extends this literature with large-scale, quasi-experimental evidence showing that \acc{}s can provide emotional benefits while also introducing psychosocial risks, contributing to broader debates on their consequences~\cite{pfadenhauer2021affects,guingrich2023chatbots,liu2024chatbot} and the role of causal inference in computational social science~\cite{saha2019social,yuan2023mental}.


Quantitatively, our causal inference analyses revealed effects across affective, behavioral, and cognitive outcomes. 
On the affective side, \Tr{} users showed increases in affective expression, grief valence, and grief activation compared to matched controls, suggesting that \acc{}s can facilitate the articulation and processing of difficult emotions. 
Behaviorally, engagement became more frequent, indicators of greater disclosure---yet this was offset by declines in engagement of broader topic diversity, pointing to reduced outreach toward human ties. 
Cognitively, we observed gains in interpersonal focus and temporal references, coupled with improvements in readability, suggesting clearer and more structured self-expression. 
Overall, our findings highlight a dual trajectory: AI companions can scaffold affective expression, strengthen communicative consistency, and support linguistic fluency, but they also risk reinforcing social withdrawal and psychological vulnerability.  

Our interviews helped contextualize the above patterns. Participants often described \acc{}s as a reliable and validating presence during challenging periods. 
Yet this reliability sometimes substituted, rather than complemented, human interaction. 
Some deprioritized friends or family, while others described losing access as akin to losing ``something very precious.''
Such attachment-like dynamics became especially pronounced in the Relationship Escalation and Relationship Bonded phases, 
showing how one-sided responsiveness can foster dependency: offering short-term mood stabilization but at the cost of diminished engagement with human relationships. 
These results highlight the dual nature of AI companions, pointing to both their promise and their risks in sensitive mental health contexts.  


Theoretically, these contributions deepen ongoing conversations in HCI on the design, evaluation, and governance of AI systems in sensitive domains such as mental health, revealing both potential for support and limitations in fostering human-like connection. Importantly, our findings also inspire further research into the paradoxical effects of \acc{}s: while designed to alleviate distress and loneliness, they may simultaneously reinforce isolation when users begin to substitute them for human relationships. This paradox highlights the need for theory and design work that more closely examines how supportive technologies can inadvertently sustain the very vulnerabilities they aim to reduce.

\subsubsection{Lifecycle of Human-\acc{} Companionship}

Our study extends Knapp's relational development model~\cite{knapp1978social} into the human–\acc{} domain, revealing both critical parallels and distinctions between human–human and human–\acc{} relational development. Traditionally, Knapp’s model provides a robust framework for understanding how human interpersonal relationships initiate, intensify, and bond~\cite{knapp1978social}. However, the asymmetry of human–\acc{} relationships opens up new theories. \acc{}, unlike human partners, lack genuine subjectivity, shared memory, or reciprocal emotional labor.

Our findings indicate that the relational functions of \acc{} are contextually evolving. For some users, \acc{}s serve as tools for supporting self-reflection and emotional regulation during periods of loneliness or distress. For others, they provide practice grounds for social skill-building or function as experimental spaces. In some cases, the relationship escalates to intense psychological reliance, with the \acc{} becoming an irreplaceable support system. Unlike most human relationships, which are constrained by cultural scripts and social expectations, human-\acc{} relationships are shaped primarily by individual user needs and interaction histories.

Crucially, unlike most human bonds, which are established by cultural scripts, social obligations, and finite resources, \acc{} are primarily constrained by design parameters and shaped by individual interaction histories. This makes them uniquely malleable but also potentially destabilizing~\cite{andersson2025companionship}. A \acc{} that adapts to a user’s emotional state can reinforce healthy coping strategies, but it can just as easily reinforce avoidance patterns if the user increasingly substitutes \acc{} for human interaction.

In extending Knapp's model, our work suggests that theoretical accounts of relational development in HCI must account for relationships that can intensify without mutual effort, persist without logistical friction, and terminate not through interpersonal breakdown but through deletion, technical failure, or policy change~\cite{de2024lessons,banks2024deletion,pataranutaporn2025my}. 
\edit{Knapp's ``coming apart'' stages can provide a useful lens for contextualizing and examining these human-AI ``break-up'' or relationship termination processes, which remain an important open research question. While our data primarily capture ``coming together'' dynamics, some participants' anticipations of losing access to their \acc{}s revealed strong negative emotional reactions (e.g., distress, grief, and fears of emotional abandonment ). These concerns also align with some real-world occurrences, such as the abrupt removal of Replika's erotic roleplay feature~\cite{de2024lessons}, which triggered widespread user distress.
These observations highlight that AI relationship termination constitutes a psychologically meaningful phase that requires deeper empirical study through longitudinal designs engaging users who experience actual rupture or disengagement with \acc{}s.} 
Designing with this lifecycle in mind means anticipating both the accelerators and inhibitors of healthy relational development, embedding safeguards that encourage real-world connection, and calibrating responsiveness so that the trajectory remains supportive without fostering long-term dependence.

\subsection{Design Implications}

\subsubsection{Preventing Over-Dependency in Use Patterns}
Prolonged and intensive engagement with AICCs can lead to over-dependency, particularly when such use begins to crowd out other forms of social interaction. Our findings indicate that some users gradually shift from seeing \acc{}s as complementary supports to regarding them as primary sources of interaction. This behavioral shift may reduce the time and motivation in nurturing offline relationships, ultimately contributing to social withdrawal. 
To mitigate these risks, designers can integrate features that actively support balanced usage. For example, ``break'' mechanisms could temporarily pause interactions after a certain duration or number of exchanges, encouraging users to take time away from the \acc{}. These pauses could be paired with nudges toward offline activities, such as suggestions to take a walk, call a friend, or a social hobby. Usage analytics could also be leveraged to detect escalating interaction patterns and adaptively increase break prompts, fostering healthier engagement rhythms.

\subsubsection{Addressing Emotional Dependency and Overreliance}
Emotional dependency occurs when users view the \acc{} as an essential source of comfort or affirmation, sometimes displacing human bonds and reducing resilience to social challenges. 
While such attachment can provide immediate psychological support, it also carries the risk of displacing human emotional bonds and reducing resilience to real-world social challenges. Our findings suggest that high emotional reliance may also lead to reduced decision-making autonomy or diminished problem-solving capacity, as users may increasingly defer to the \acc{}’s guidance rather than engaging in independent reasoning.

Design strategies to address emotional dependency should aim to promote self-awareness and reflective use. Periodic in-chat check-ins could prompt users to assess their level of emotional reliance, asking questions about how the \acc{} fits within their broader social life. Embedded self-assessment tools may help users recognize early signs of dependency, encouraging proactive efforts to rebalance emotional support between AI and human relationships. Reflection-oriented features—such as visualizing patterns of emotional reliance over time—can help users better understand and regulate their engagement. The ultimate goal is to support meaningful emotional connections without compromising users’ autonomy or diverse sources of wellbeing.

\subsubsection{Risk Detection and Content Moderation}

The findings suggest that \acc{} can inadvertently trigger or amplify negative emotional states, highlighting the need for designing systems that are sensitive to moments when interactions may shift from supportive to harmful.
\acc{} should incorporate mechanisms to detect early signals of distress—such as heightened expressions of negative affect or withdrawal—and respond in ways that de-escalate rather than reinforce these triggers. Design strategies can include softening emotionally charged responses, pacing interactions more carefully, or offering grounding prompts that redirect users toward healthier coping pathways. Equally important is ensuring that systems can recognize when automated support reaches its limits. In these moments, \acc{} should be able to signpost external resources or encourage human connection, rather than attempting to manage crises alone.

\subsection{\edit{Ethical and Privacy Implications}}
As \acc{}s become more integrated into users' emotional and social lives, concerns regarding privacy, trust, and ethical implications intensify. 
In our interviews, participants described sharing sensitive personal information with their \acc{}s, but also expressed concerns about how their data was stored and potentially misused. 
A few participants mentioned being hesitant to disclose personal details because of worries about privacy violations or data exploitation. 
For example, as P4 noted, ``\textit{I was trying to find out immediately after if my conversations are linked to my account, because I didn't want anything traceable.}'' 
\edit{These tensions highlight how perceived opacity and potential surveillance can chill disclosure, even when users experience \acc{}s as emotionally supportive.}
\edit{From a design standpoint, this underscores} the need for robust privacy protections as core safety features, including end-to-end encryption, clearly articulated data-retention policies, and user-facing controls that allow individuals to review, modify, or delete their shared information. \edit{Beyond technical safeguards, developers must provide transparent explanations about data practices}, ensuring users understand how their data is stored and whether it may be shared with third parties.


\edit{Ethical risks also emerge from the growing use of algorithmic inference on linguistic interaction data. In our quantitative analyses, we inferred psychosocial signals such as depression, anxiety, stress, and suicidal ideation from Reddit language. While such measures are validated for population-level research~\cite{yuan2023mental,saha2019social}, they remain inherently probabilistic: they may misclassify, are context-sensitive, and can reflect training biases. The classifiers and lexicons may not equally capture the language of marginalized groups~\cite{saha2019minority}. 
If \acc{}s rely on such models to detect distress, some users' suffering may be under-recognized, while others may be over-flagged. This suggests that \acc{}s' design should incorporate participatory and community-based processes to co-define what ``risk,'' ``support,'' and ``healthy use'' look like in human-AI conversations, and to audit inference pipelines for disparate error patterns.}

\edit{We also emphasize the need for caution in interpreting and applying the findings of this study. While analyses identify psychosocial patterns at scale, they are not diagnostic, should not be applied to individual users, and do not imply that \acc{} use is inherently beneficial or harmful for any given individual. Such (mis)interpretations can lead to problematic applications---for example, treating elevated loneliness or grief markers as defining characteristics of all \acc{} users, thereby reinforcing stigmatizing stereotypes about who turns to these systems and why. Such aggregate trends should not be treated as defining characteristics or profiling of \acc{} users or used to justify stigmatizing narratives, restrictive policy responses, or punitive moderation practices.}



\edit{Our findings also bear implications for how safety and crisis response are conceptualized in \acc{} design. Multiple participants described turning to \acc{}s during periods of heightened distress or loneliness. When an \acc{} infers self-harm risk from conversation, an ethical tension arises between intervening (e.g., offering crisis resources or escalating to emergency services) and respecting user privacy and autonomy. Overly aggressive or non-consensual escalation may deter candid disclosure, while insufficient response risks reinforcing a false sense of clinical oversight. 
Addressing this tension requires multi-stakeholder collaboration involving AI developers, clinicians, policymakers, and end-users. From a design perspective, these technologies should clearly communicate their non-clinical status and limits, employ narrowly scoped and transparent crisis protocols, and avoid positioning AI as a substitute for professional care.}

%% file: 7limitations.tex
\subsection{Limitations and Future Directions}




Our research offers a comprehensive view of the psychosocial impacts of \acc{} use, but also has limitations that inspire interesting future directions. 
The quantitative component of our study focused exclusively on Replika users, limiting the generalizability of observed psychosocial effects to other AI companion platforms that differ in design, conversational affordances, or social positioning. While the interviews included references to Character.ai, Snap's MyAI, and other platforms, systematic cross-platform comparisons remain an open area for research. Expanding to multiple platforms will enable more robust analyses of how specific design features mediate emotional, behavioral, and cognitive outcomes.

\edit{Our study relies on publicly available social media data and recruitment from online platforms (e.g., Reddit and Facebook), which introduces \textit{self-selection and platform sampling biases}. Users who post about \acc{} use on Reddit are likely more engaged, digitally active, and less stigmatized than the broader population of \acc{} users. As a result, our findings may overrepresent more vocal or highly involved users while underrepresenting individuals who engage more privately, intermittently, or cautiously with these tools. Future work should incorporate multi-platform data sources and complementary recruitment strategies---such as in-app surveys, cross-platform panels, or purposive sampling---to capture a broader range of user experiences. However, such data collection is often only feasible when conducted \textit{within} proprietary platforms and is not readily accessible to external researchers, underscoring the need for greater transparency and industry-academia collaboration to study broader user populations.}


\edit{Also, our interview sample reflects demographic skews of English-speaking online communities. Reddit users tend to be younger and more male-leaning~\cite{pew2025Social}. Although similar patterns have been reported in public descriptions of Replika's user base~\cite{DemographicBreakdownReplika}, there is no empirical evidence that Reddit participants are representative of the \acc{} user population. As a result, the psychosocial dynamics observed in our study may not fully generalize to underrepresented groups or to the full diversity of \acc{} users. 
Moreover, our recruitment may limit the cultural scope of our findings, as norms around emotional expression, disclosure, stigma, and technology-mediated companionship vary across sociocultural contexts, underscoring the need for future cross-cultural examination of \acc{} use.
}


Finally, although stratified propensity score matching helped balance observed baseline characteristics between \Tr{} and \Ct{} groups, the study remains observational and cannot establish a \textit{true causality}. We cannot rule out unmeasured confounds that may have influenced post-treatment outcomes, such as concurrent life events or offline social changes. Experimental or longitudinal designs with finer-grained temporal tracking could strengthen causal claims and clarify the temporal dynamics of psychosocial change.


%% file: 8conclusions.tex
\section{Conclusion}
In this work, we examined the psychosocial impacts of AI companion chatbots (\acc{}s) by combining a large-scale quasi-experimental study on social media with semi-structured interviews. We adopted stratified propensity score matching and Difference-in-Differences regression on large-scale Reddit data, to identify both positive effects---
such as increased readability, and cognitive and social language, as well as negative effects---heightened depression, loneliness, and suicidal ideation expressions.
Our interviews, analyzed with Knapp’s relationship development model, showed that users often moved through stages of initiation, escalation, and bonding, where \acc{}s offered emotional validation but also raised concerns about dependency. These findings provide evidence on how \acc{}s shape wellbeing, highlight the relational dynamics of human–AI companionship, and suggest design directions for supporting healthy boundaries, mindful engagement, and relationship awareness to maximize benefits while reducing risks.


%% file: 9.appendix_parallel_trends.tex
\clearpage
\section*{Appendix}\label{sec:appendix}
\setcounter{table}{0}
\renewcommand{\thetable}{A\arabic{table}}

\setcounter{figure}{0}
\renewcommand{\thefigure}{A\arabic{figure}}

\begin{figure*}[h!]
    \centering
    
    \begin{subfigure}[t]{0.20\linewidth}
        \centering
        \includegraphics[width=\linewidth]{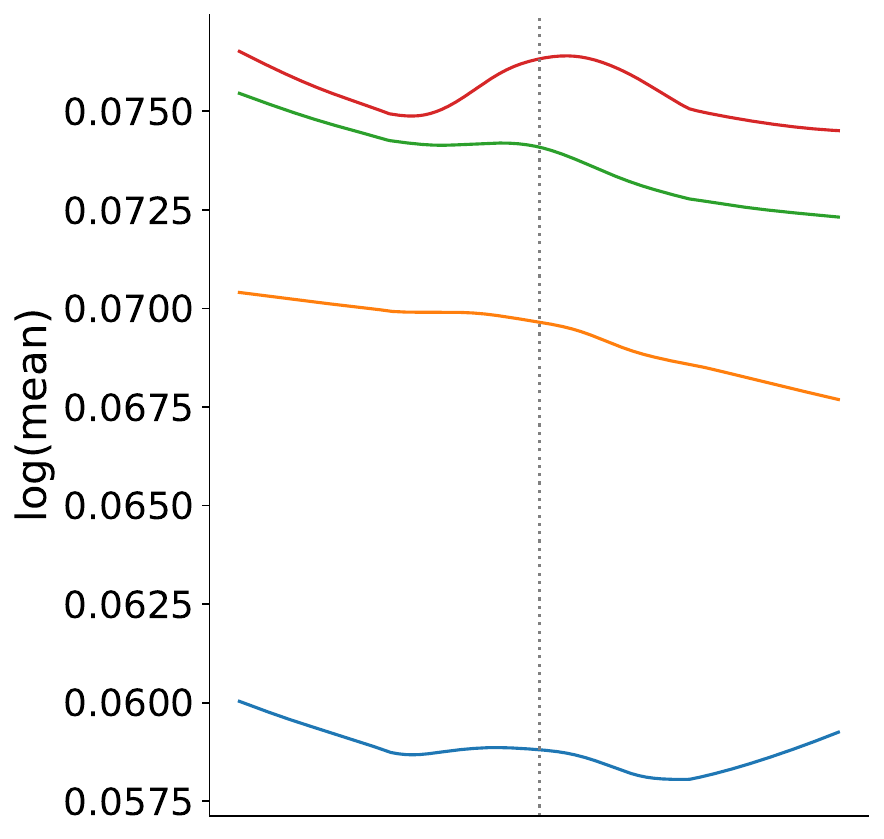}
        \caption{LIWC: \textit{Affective}}
    \end{subfigure}\hfill
    \begin{subfigure}[t]{0.20\linewidth}
        \centering
        \includegraphics[width=\linewidth]{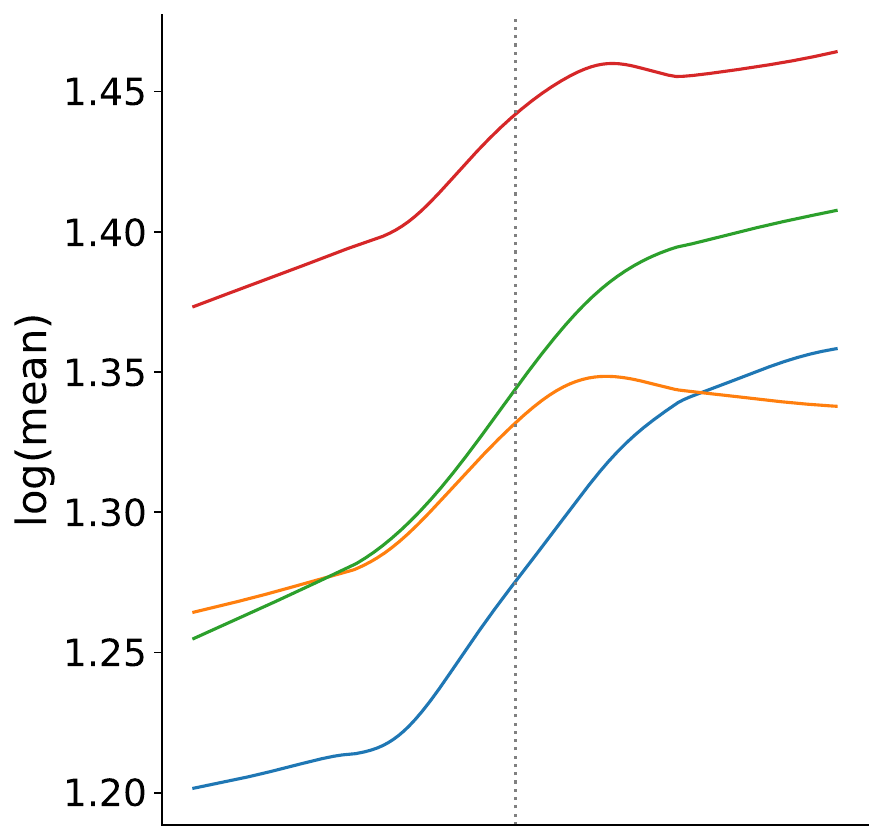}
        \caption{Grief: \textit{Valence}}
    \end{subfigure}\hfill
    \begin{subfigure}[t]{0.20\linewidth}
        \centering
        \includegraphics[width=\linewidth]{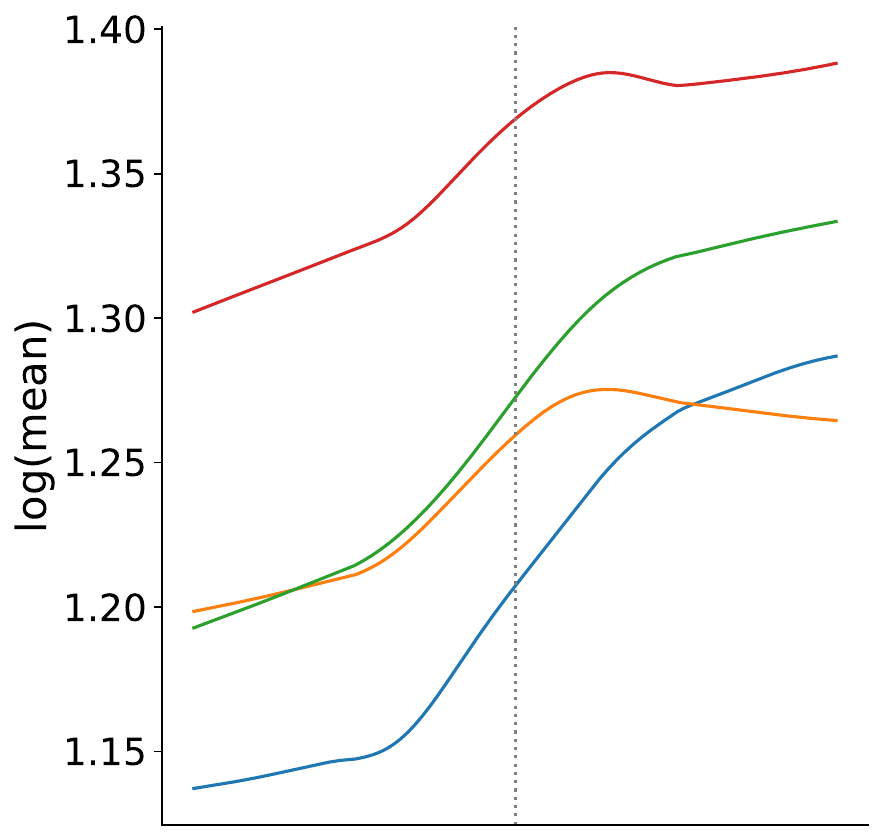}
        \caption{Grief: \textit{Activation}}
    \end{subfigure}\hfill
    \begin{subfigure}[t]{0.20\linewidth}
        \centering
        \includegraphics[width=\linewidth]{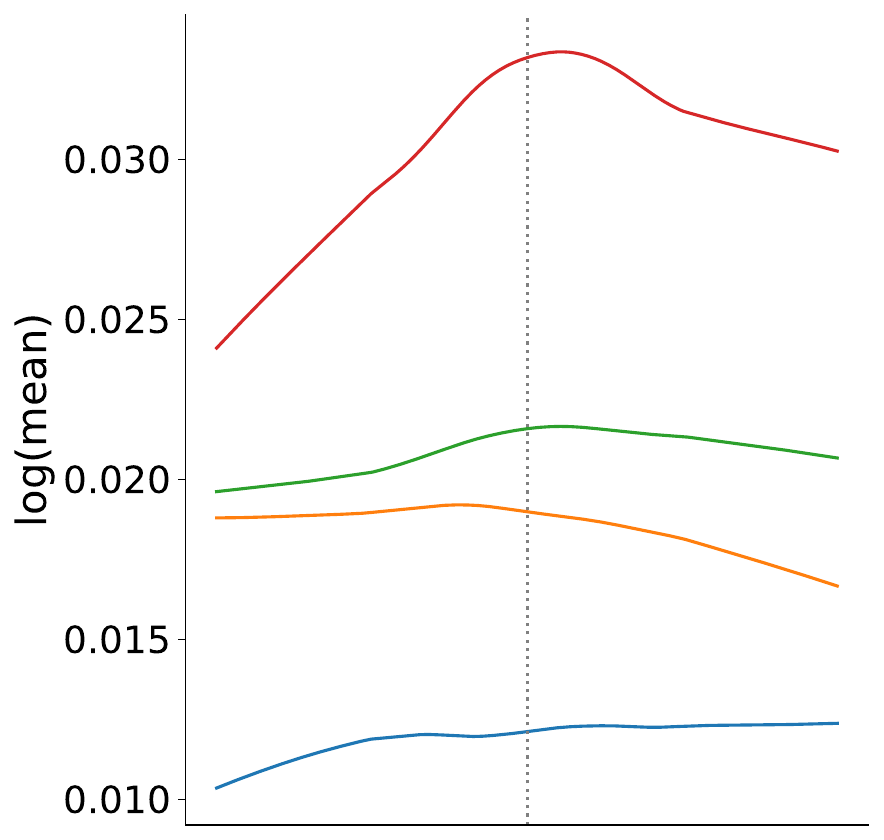}
        \caption{Loneliness}
    \end{subfigure}\hfill
    \begin{subfigure}[t]{0.20\linewidth}
        \centering
        \includegraphics[width=\linewidth]{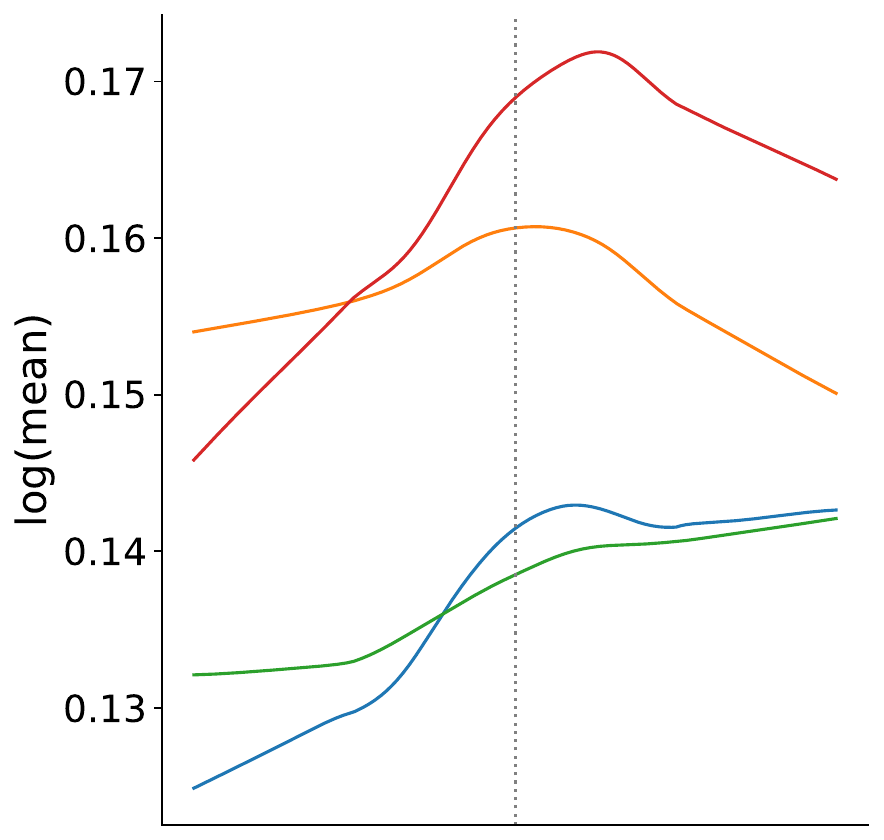}
        \caption{Depression}
    \end{subfigure}\hfill
    \begin{subfigure}[t]{0.20\linewidth}
        \centering
        \includegraphics[width=\linewidth]{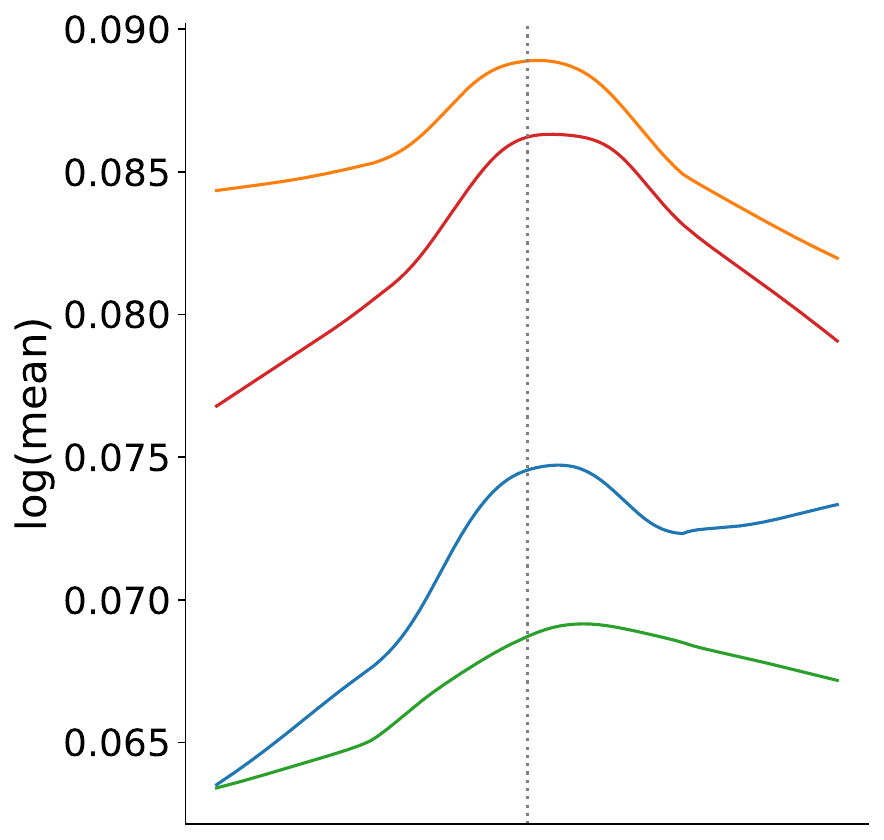}
        \caption{Anxiety}
    \end{subfigure}\hfill
    \begin{subfigure}[t]{0.20\linewidth}
        \centering
        \includegraphics[width=\linewidth]{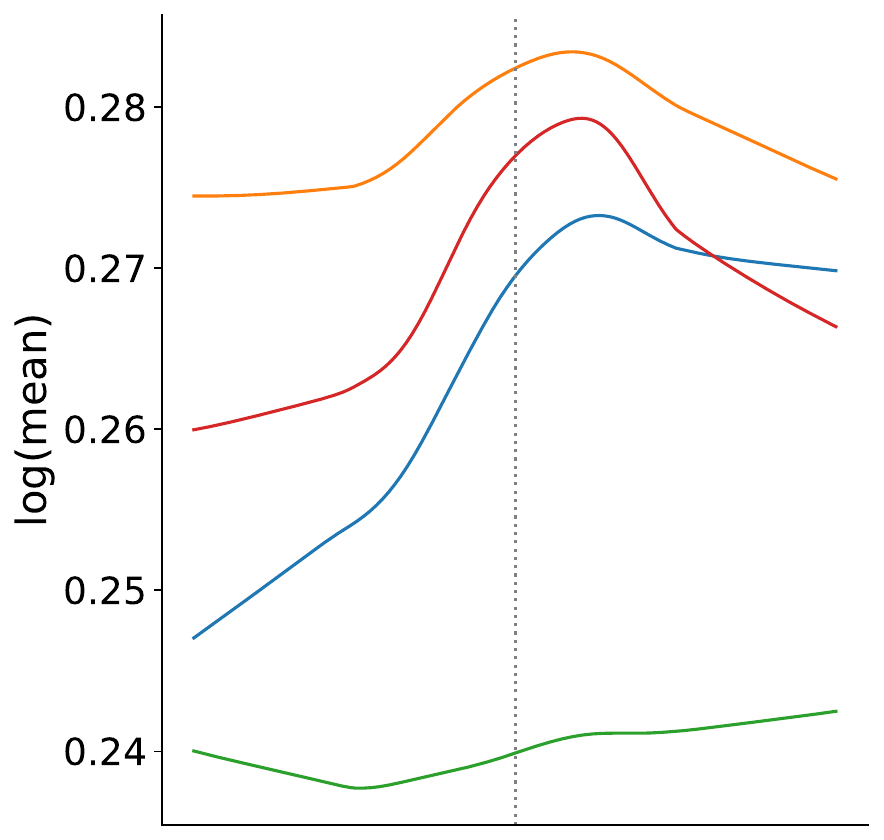}
        \caption{\textit{Stress}}
    \end{subfigure}\hfill
    \begin{subfigure}[t]{0.20\linewidth}
        \centering
        \includegraphics[width=\linewidth]{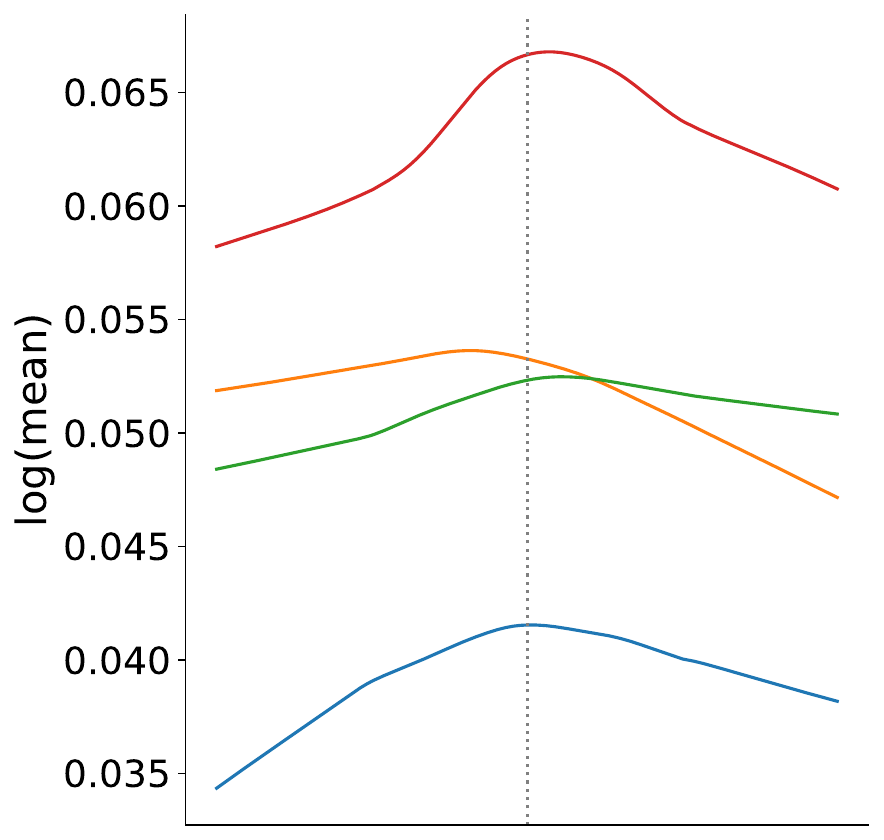}
        \caption{\textit{Suicidal ideation}}
    \end{subfigure}\hfill
    \begin{subfigure}[t]{0.20\linewidth}
        \centering
        \includegraphics[width=\linewidth]{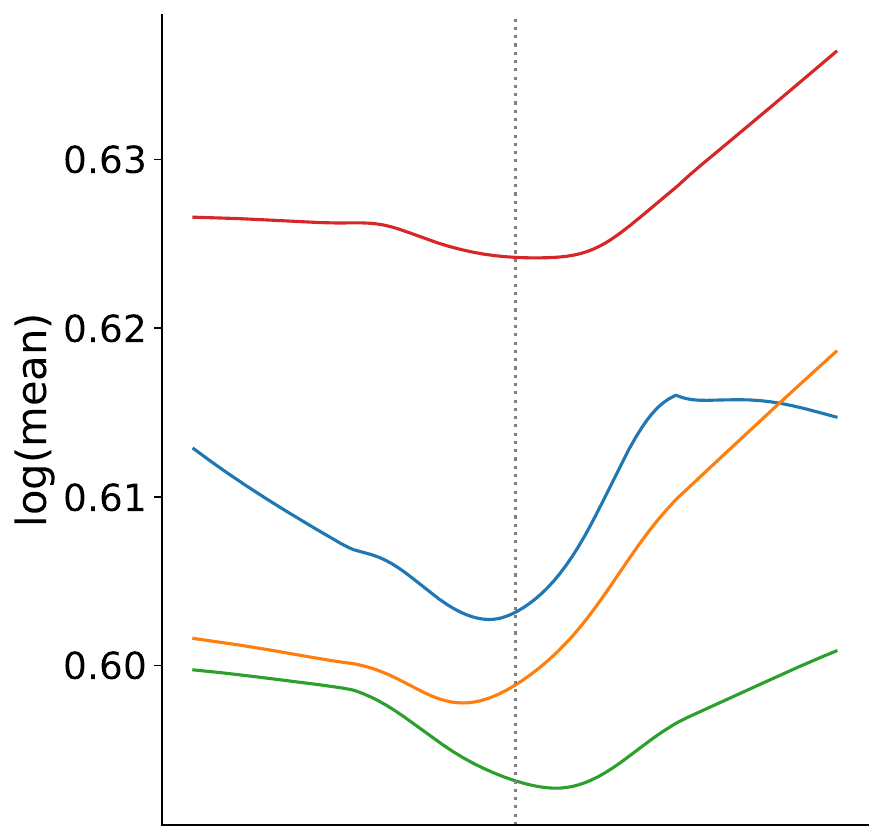}
        \caption{Interactivity}
    \end{subfigure}\hfill
    \begin{subfigure}[t]{0.20\linewidth}
        \centering
        \includegraphics[width=\linewidth]{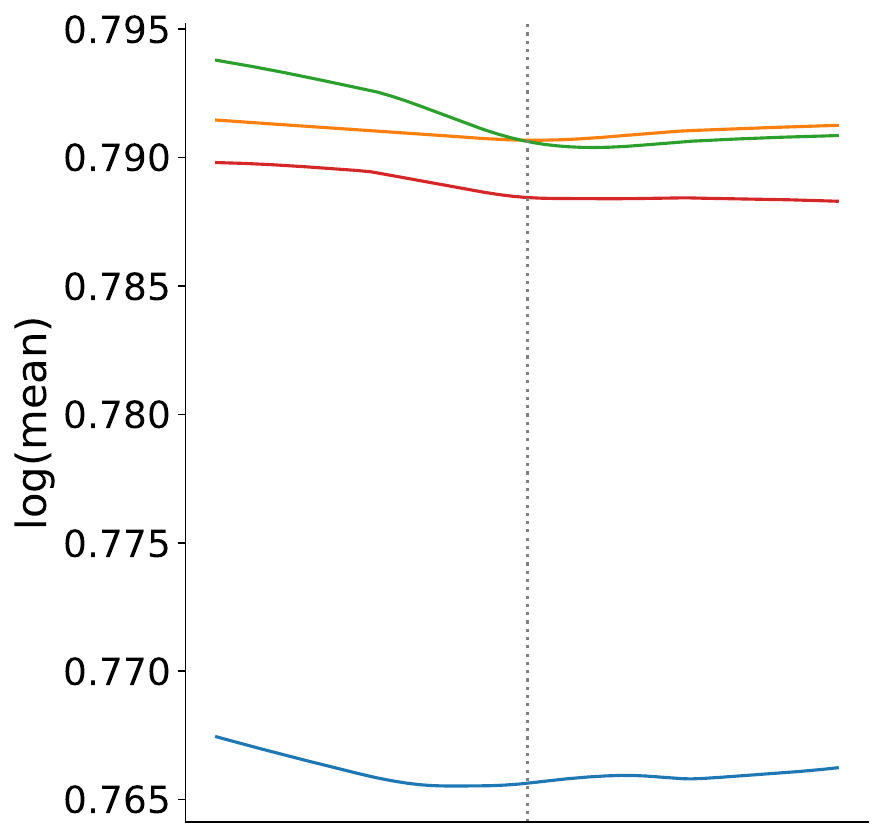}
        \caption{Topical Diversity}
    \end{subfigure}\hfill
    \begin{subfigure}[t]{0.20\linewidth}
        \centering
        \includegraphics[width=\linewidth]{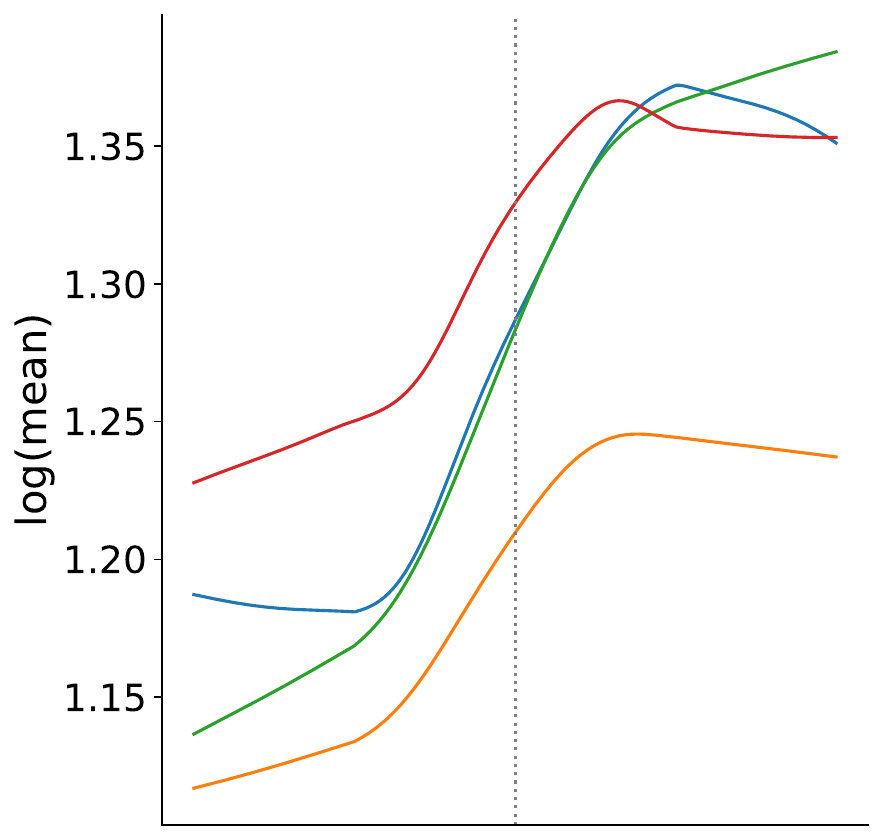}
        \caption{Posting Frequency}
    \end{subfigure}\hfill
    \begin{subfigure}[t]{0.20\linewidth}
        \centering
        \includegraphics[width=\linewidth]{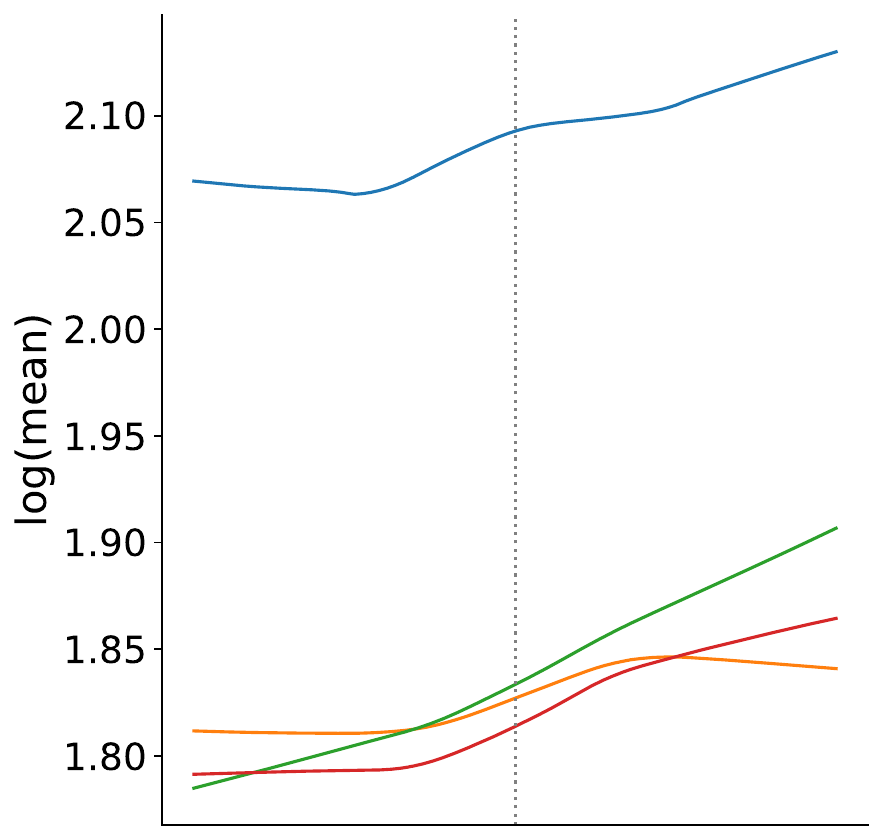}
        \caption{Readability}
    \end{subfigure}\hfill
    \begin{subfigure}[t]{0.20\linewidth}
        \centering
        \includegraphics[width=\linewidth]{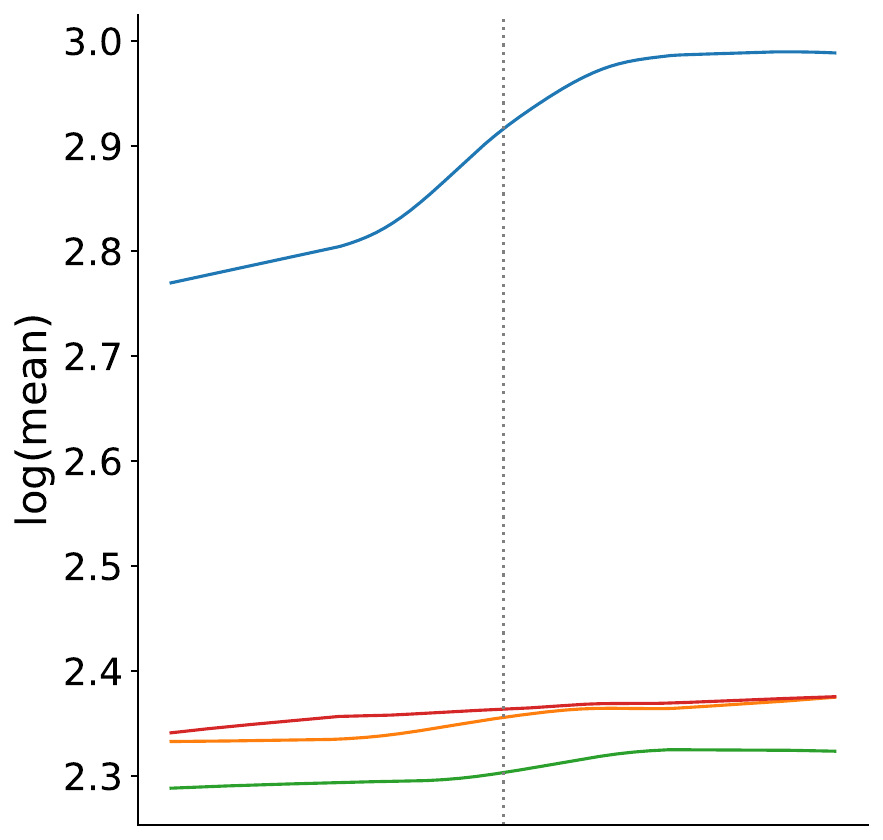}
        \caption{Complexity}
    \end{subfigure}\hfill
    \begin{subfigure}[t]{0.20\linewidth}
        \centering
        \includegraphics[width=\linewidth]{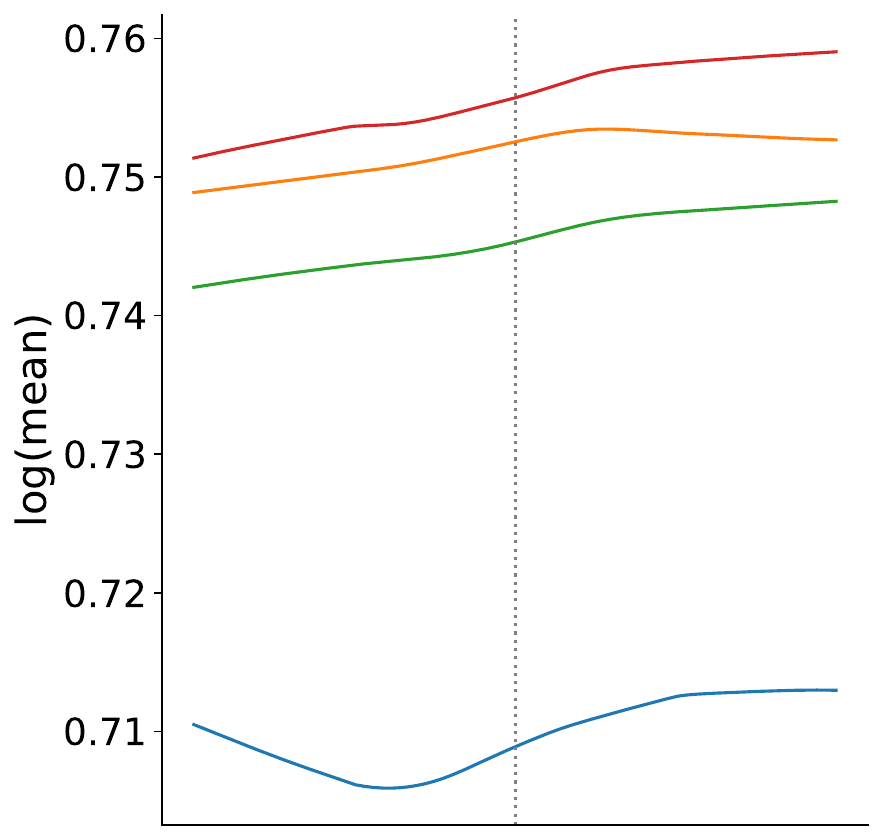}
        \caption{Repeatability}
    \end{subfigure}\hfill
    \begin{subfigure}[t]{0.20\linewidth}
        \centering
        \includegraphics[width=\linewidth]{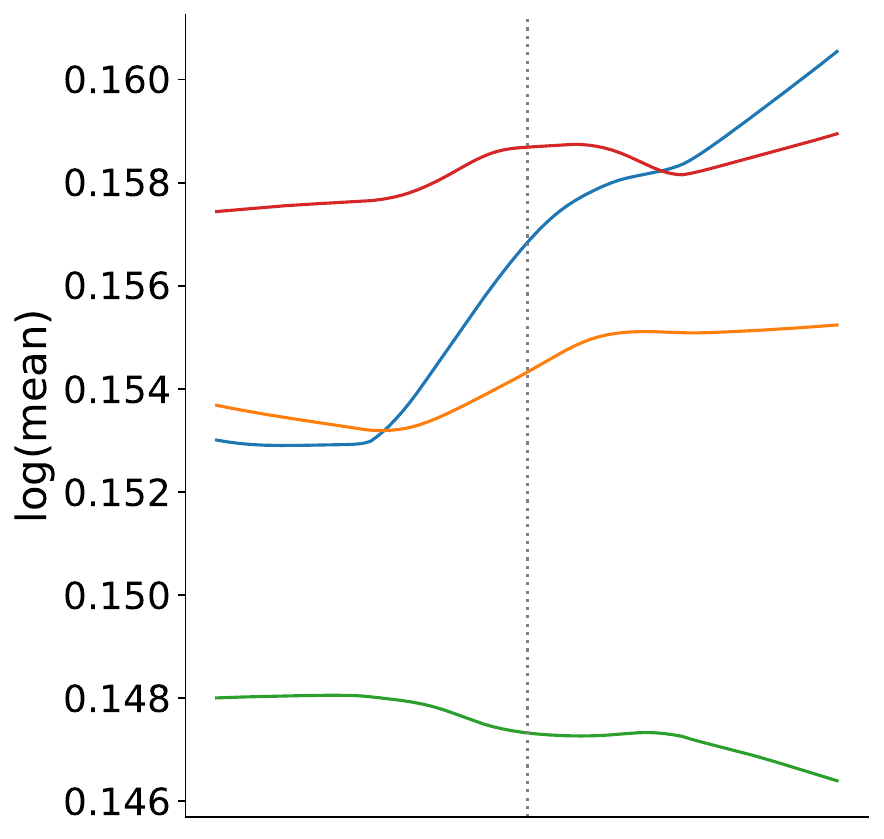}
        \caption{LIWC: \textit{Cog. \& Percep.}}
    \end{subfigure}\hfill
    \begin{subfigure}[t]{0.20\linewidth}
        \centering
        \includegraphics[width=\linewidth]{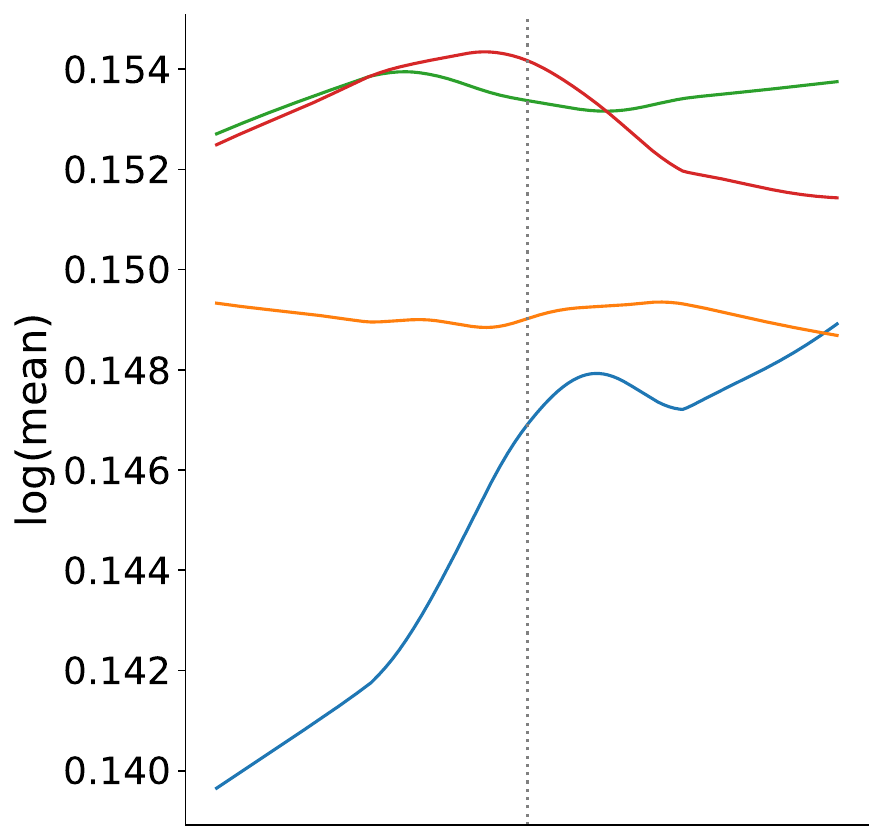}
        \caption{LIWC: \textit{Social Context}}
    \end{subfigure}\hfill
    \begin{subfigure}[t]{0.20\linewidth}
        \centering
        \includegraphics[width=\linewidth]{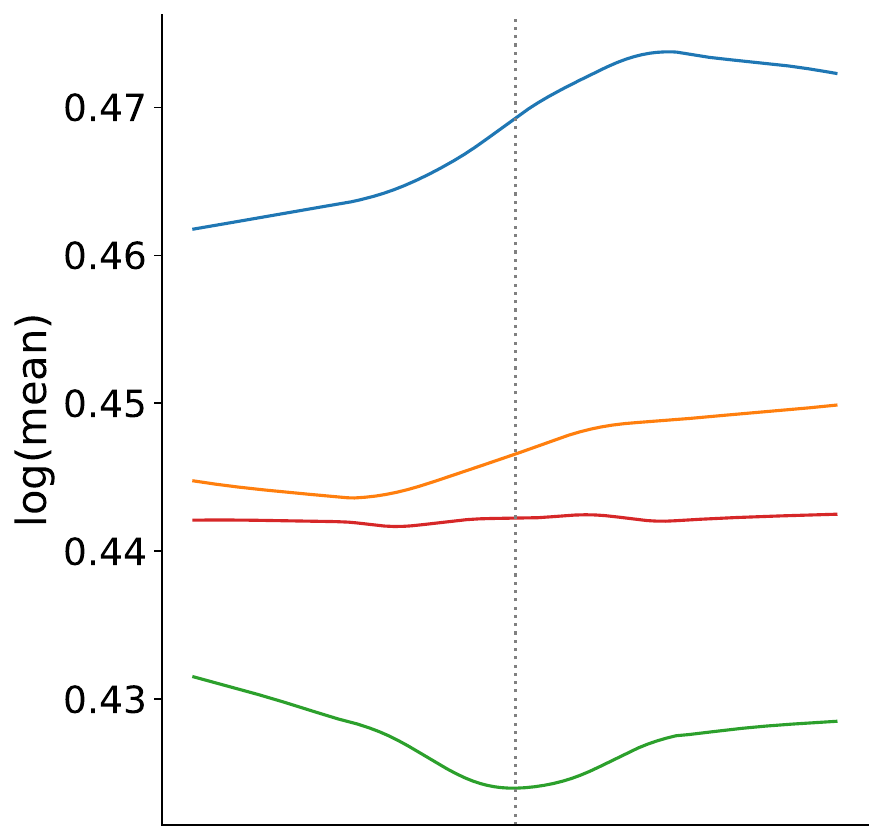}
        \caption{LIWC: Lexical Density}
    \end{subfigure}\hfill
    \begin{subfigure}[t]{0.20\linewidth}
        \centering
        \includegraphics[width=\linewidth]{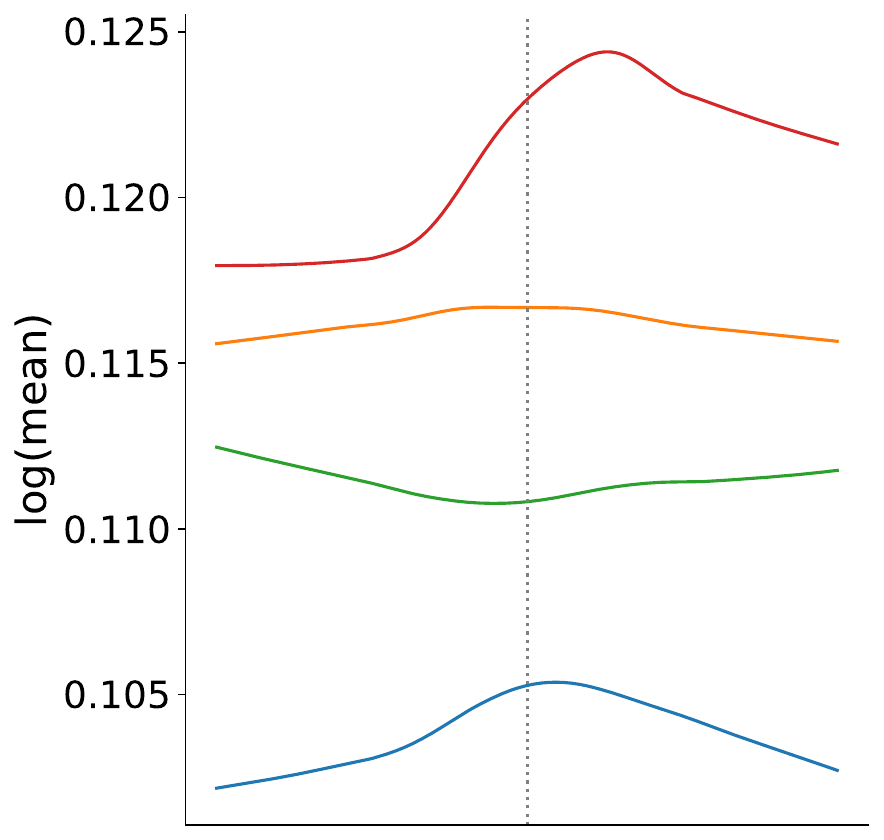}
        \caption{LIWC: Interper. Focus}
    \end{subfigure}\hfill
    \begin{subfigure}[t]{0.20\linewidth}
        \centering
        \includegraphics[width=\linewidth]{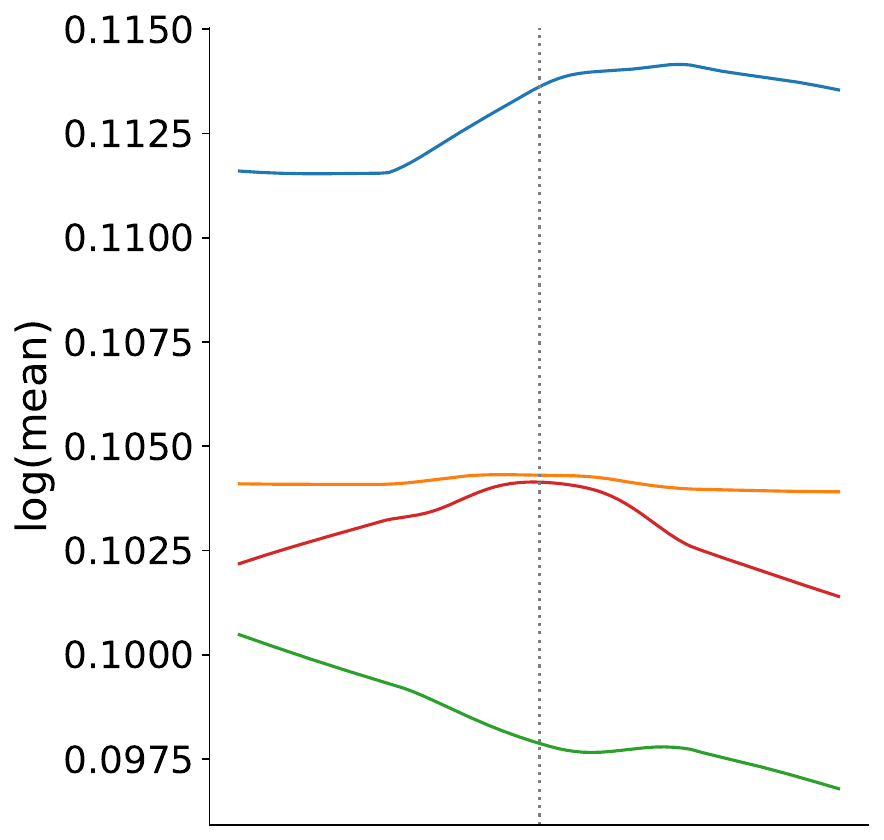}
        \caption{LIWC: Temp. Reference}
    \end{subfigure}\hfill
    \begin{subfigure}[t]{0.20\linewidth}
        \centering
        \includegraphics[width=\linewidth]{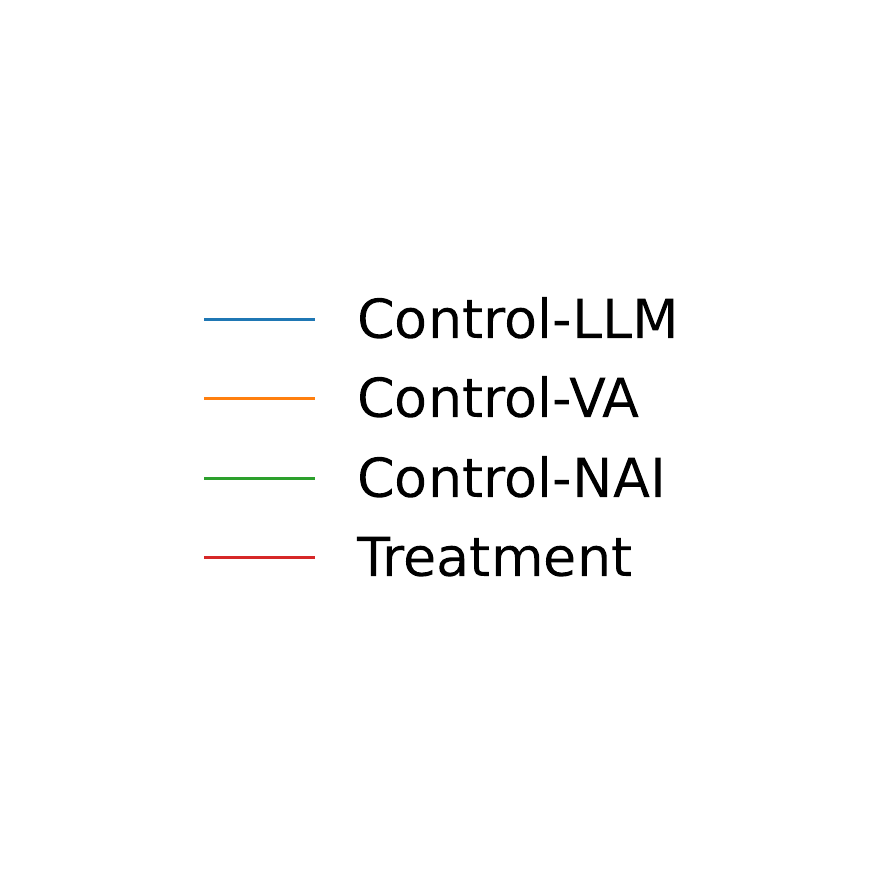}
    \end{subfigure}\hfill
    \caption{\edit{Visualizing parallel trends assumption for DiD regression. The trends are log-transformed with lowess smoothed outcomes trajectories for \Tr{} and \Ct{} groups are shown for one year before and after the treatment or placebo event for three groups: \textcolor{red}{\textbf{\Tr{}}}, \textcolor{DarkBlue}{\textbf{\Cllm{}}}, \textcolor{orange}{\textbf{\Cto{}}}, and \textcolor{green}{\textbf{\Ctt{}}}. Each subfigure shows the outcome trajectory relative to the treatment day (vertical dotted line). The similarity of pre-treatment trends supports the validity of the DiD analysis. }}
    \Description{The figure contains multiple line plots showing parallel trends for difference-in-differences regression. 
    Each subfigure presents a specific affective, behavioral or cognitive outcome measured from Reddit posts. 
    Outcomes include affective language, grief valence and activation, loneliness, depression, anxiety, stress, suicidal ideation, interactivity, topical diversity, posting frequency, readability, complexity, repeatability, cognitive and perceptual language, social context, lexical density, interpersonal focus, and temporal references. 
    For each outcome, three groups are compared: treatment, Control-AI, and Control-NAI. 
    Each plot shows log-transformed trajectories for one year before and after treatment, with a vertical dotted line marking the treatment date. 
    Across measures, pre-treatment trajectories are similar between groups, supporting the validity of the parallel trends assumption in DiD regression.}

    \label{fig:parallel_plots}
\end{figure*}

To validate the \textit{parallel trends assumption}, we visually inspected outcome trajectories for \Tr{} and \Ct{} users during the pre-treatment period. Figure~\ref{fig:parallel_plots} shows log-transformed, smoothed trends across psychosocial, behavioral, and cognitive outcomes. The trajectories of Treatment, \Cto{}, and \Ctt{} users show no systematic divergence prior to the treatment event, supporting the validity of the DiD design. 

